\documentclass[aps,titlepage,12pt]{revtex4}
\usepackage{amsfonts}
\usepackage{amsmath}
\usepackage{graphicx}
\usepackage{dcolumn}
\usepackage{bm}
\usepackage{overpic}
\usepackage{booktabs}
\usepackage{color}
\usepackage[justification=raggedright,font={small,sf}, singlelinecheck=false]{caption}

\begin{document}
\title{The critical role of fresh teams in creating original and multi-disciplinary research}

\author{An Zeng$^1$, Ying Fan$^1$, Zengru Di$^1$, Yougui Wang$^1$, Shlomo Havlin$^{2,}\footnote{havlin@ophir.ph.biu.ac.il}$}

\affiliation{
$^1$ School of Systems Science, Beijing Normal University, Beijing, China.\\
$^2$ Department of Physics, Bar-Ilan University, Ramat-Gan 52900, Israel.
}

\begin{abstract}
Teamwork is one of the most prominent features in modern science. It is now well-understood that the team size is an important factor that affects team creativity. However, the crucial question of how the character of research studies is influenced by the freshness of the team remains unclear. In this paper, we quantify the team freshness according to the absent of prior collaboration among team members. Our results suggest that fresher teams tend to produce works of higher originality and more multi-disciplinary impact. These effects are even magnified in larger teams. Furthermore, we find that freshness defined by new team members in a paper is a more effective indicator of research originality and multi-disciplinarity compared to freshness defined by new collaboration relations among team members. Finally, we show that career freshness of members also plays an important role in increasing the originality and multi-disciplinarity of produced papers.
\end{abstract}

\date{\today}

\maketitle

\section{Introduction}
In contrast to about a century ago, when individual scientists played a significant role in scientific discoveries, teamwork is becoming
increasingly common in recent modern science~\cite{science2018fortunato,the2017zeng}. Indeed, it has been found that the fraction
of scientific papers that were written by teams and the mean team size increased during the last century, indicating a significant shift
towards a teamwork~\cite{the2007wuchty,team2005guimera,from2016leahey}. Specifically, it has been shown that the mean team size of research papers increased from 1.9 to 3.5 authors per paper from 1955 to 2000~\cite{the2007wuchty}. Apart from that, the team size distribution has been found to change fundamentally from a simple Poisson distribution to a power-law shape distribution~\cite{principles2014milojevic}. These phenomena are attributed to the combination effect of the increasing scale, complexity, and costs of big science~\cite{collaborative2008hunter,inequalities2014xie,mapping2011falk}.

Various models have been developed to better understand team formation in scientific research. A team is defined as the coauthors of a paper and many classical works focus on studying the collaboration features of individual scientists in order to understand team formation~\cite{evolution2002barabasi}. Related studies are numerous, mainly aiming to reveal the topological features such as community structure and assortative mixing in collaboration networks~\cite{the2001newman,quantify2015petersen} and model the evolution of collaboration networks and author-paper bipartite networks~\cite{evolve2007li,the2004borner}. In recent years, attention has been shifted to directly understand the team assembly mechanisms. For example, a recent work found that research teams include both small, stable ``core" teams and large, dynamically changing ``extended" teams~\cite{principles2014milojevic}. The shift of team size distribution from Poisson to power-law has been explained by the fast tendency towards extended teams~\cite{principles2014milojevic}. Another study investigates how the mechanisms by which creative teams self-assemble determine the structure of collaboration networks, and observe a second-order phase transition of the giant component in the collaboration networks~\cite{team2005guimera}.

In many studies, the citations of papers have been used to measure the impact of the paper~\cite{how1998redner}. By comparing papers of multiple
authors to papers that have a single author, a strong signal favoring teamwork has been detected~\cite{the2007wuchty}. The distribution of workload across team members is shown to largely affect the performance of teams~\cite{understand2016klug}. It is also found that more authors and countries in a paper are associated with higher citation rates when examining the influence of international research teams on citation outcomes~\cite{multinational2015hsiehchen}. In a recent work, the authors use an index called disruption to measure the originality of a paper~\cite{large2019wu}. They interestingly find that small teams tend to disrupt science and technology with original ideas and opportunities while larger teams tend to develop existing ones. This finding highlights the vital role that small teams have in expanding the frontiers of knowledge.

So far, various factors such as team size~\cite{the2007wuchty,large2019wu}, workload distribution~\cite{understand2016klug}, number of involved countries~\cite{multinational2015hsiehchen,evolution2016coccia}, universities~\cite{multi2008jones,mapping2012gazni} and disciplines~\cite{inter2015van,aytpical2013uzzi} have been found to remarkably affect the outcome impact of teamwork. However, the role of team freshness in advancing science has been rarely studied. A research team may consist of some researchers who have not worked with each other before, resulting in some freshness of the team. On the contrary, authors in a paper may have already published a number of papers together, thus working as an old team. In addition, less seniority of team members in their careers can be also regarded as some freshness of a team. In team formation models, the tendency of individual teams in selecting new team members has been interestingly found to decrease the giant component of the whole collaboration network and this tendency varies in papers of different journals~\cite{team2005guimera}. However, how the team freshness affects the performance of teams in advancing science remains unclear.

In this study, we address the effect of team freshness on the originality and multidisciplinary of the produced work, by investigating systematically the prior collaboration relations between team members. The freshness of a team is defined according to the fraction of team members that have not been collaborated earlier with other team members, see Fig. 1. We find that papers of fresher teams are significantly more effective than papers of older teams in creating studies of higher originality and more multi-disciplinary impact. We find that the effect is even more prominent in larger teams. Our results suggest that having new team members is more powerful than new collaboration relations in increasing the originality and impact diversity of the resultant papers. Finally, we also study the effect of the career freshness of team members and find that the younger is the team, the higher are the originality and impact diversity of the produced studies.

\section{Results}
We begin by defining the freshness of a team of a paper as the fraction of team members who have not collaborated with any of other team members before they coauthor this paper. According to this definition, the freshness varies between $0$ and $1$, corresponding to fully old teams and fully fresh teams, respectively. This definition can be easily calculated by constructing a collaboration network representing all prior collaboration relations among the team members of the considered paper. The freshness of a team can be obtained directly by computing the fraction of nodes with zero degree in this collaboration network. The definition is illustrated in Fig. 1, with two networks describing examples of freshness $0$ and $1$.

In this paper, we analyze the scientific publication data of the American Physical Society (APS) journals, containing 482,566 papers, ranging from year 1893 to year 2010. Evaluating the freshness of a team requires the knowledge of the prior papers that each team member published. Thus, we need to assign each paper in the data set correctly to its real authors. For this we have used the disambiguated author name data provided in~\cite{quantifying2016sinatra} to assign each paper to its authors, which results in 236,884 distinct scientists and 482,566 papers. Furthermore, we also examine two additional data sets from computer science and chemistry. Since their results are consistent with those on the APS data, the analyses in the main paper are based on the APS data while the results of the other two data sets are presented in Fig. S12-S15 of the supplementary materials (SM). In Figs. 1c to 1e, we show the distribution of team freshness for 2-author papers, 4-author papers and 8-author papers, respectively. The results are consistent with the intuition that fresh teams are more common in small teams than in large teams. The teams with freshness $1$ are $54\%$ of all 2 author teams, while freshness $1$ fraction is only $4.6\%$ for 8-author teams.

To evaluate the role that team freshness has in advancing science, we consider a recently developed index~\cite{large2019wu,dynamic2017funk}, disruption $D$, to measure the originality of the resultant study, and propose here a measure for multi-disciplinary impact, $M$, to evaluate the diversity of disciplines that a paper influences (see Fig. 1 for illustration and Methods for more details on both measures). The disruption, $D$, varies between $-1$ and $1$. A larger disruption of a paper reflects that more of the paper's citing papers cite it but none of its references, corresponding to higher originality. The multi-disciplinary impact, $M$, of a paper is defined as the fraction of its temporal adjacent citing papers that share no other reference apart from the focal paper. The multi-disciplinary impact varies between $0$ and $1$, corresponding to narrow and diverse impact in different disciplines, respectively.

The first question we ask here is whether and how team freshness affects the originality and impact diversity of the produced work. To answer this we show in Fig. 2 the disruption $D$ representing originality and multi-disciplinarity $M$ of papers as a function of different team freshness. Fig. 2 contains results for 2-author papers, 4-author papers, and 8-author papers. The results for all cases exhibit a consistent increasing trend of both originality (disruption) and multi-disciplinarity with increasing team freshness. To examine the significance of the trend, we compare the distributions of the bootstrap disruption and bootstrap multi-disciplinarity of papers with team freshness 0 and 1 (for details see Methods). The results are presented in the inset of the panels in Fig. 2. A remarkable difference of the distributions can be observed between papers with team freshness 0 and 1. To further support the significance of the trend, we test differences in the distribution of disruption between all different team freshness using two-sample Kolmogorov-Smirnov tests. The results shown in SM Fig. S1(a-c), suggest that even with small difference in team freshness, the difference in originality (disruption) is significant with most $p$-values being smaller than 0.01. Similar results have been obtained for the multi-disciplinary impact trend using the Kolmogorov-Smirnov test, see Fig. S1(d-f).

One possible concern regarding the observed trend in Fig. 2 is whether the disruption and multi-disciplinary impact detects the same property of a paper, such that the increasing trend of one index with team freshness is highly related to the increase of the other index. To test this, we first study the relation between these two indexes in SM Fig. S2. We find (a) that both indexes, $D$ and $M$, have no correlation with citations and (b) the Pearson correlation coefficient between disruption and multi-disciplinary impact is 0.32, indicating some correlations. Next, we examine how much both indices are independent. To this end we study the relation between disruption and team freshness when controlling the multi-disciplinary impact. In Fig. S3 of SM, we respectively analyze papers with multi-disciplinary impact $M\approx0.3$, $M\approx0.5$ and $M\approx0.7$, and find that even for fixed $M$ the disruption of these papers still increases with team freshness. Similarly, we fix disruption to be $D\approx-0.1$, $D\approx0$ and $D\approx0.1$, and find that even for fixed $D$ the multi-disciplinary impact still increases with team freshness, respectively. These results suggest that the disruption and multi-disciplinary impact truly represent quite distinct properties.

The team members in fresh teams, according to our definition, do not have prior collaboration with any other in the team. The team freshness might actually to some degree be related to the prior productivity of team members. If team members have fewer papers before, the formed team is more likely to be a fresh team. It is thus important to test whether the observed trend with freshness in Fig. 2 can be simply explained by the prior productivity of team members. To remove this effect we study in Fig. S4 of SM the dependence of disruption and multi-disciplinary impact on freshness when controlling the team member productivity. The results suggest that the increasing trend of disruption and multi-disciplinary impact with team freshness is preserved, indicating that team freshness indeed play a critical role in affecting the originality and impact diversity of the produced works. As the APS data ranges from year 1893 to year 2010, we also examine the mean disruption (originality) and mean multi-disciplinary impact of papers in different years. We show in Fig. S5 of SM that both indexes decrease with time yet fresh teams constantly have higher originality and multi-disciplinarity than old teams.

Team size has been found to be an important factor in affecting the disruption of a paper, i,e., the disruption (originality) decreases with team size~\cite{large2019wu}. It is therefore natural to ask how team freshness affects disruption and multi-disciplinarity in teams of different sizes. To this end, we analyze in Fig. 3 the mean disruption $D$ and multi-disciplinary impact $M$ as a function of team size of papers published by old teams (freshness=0) and fresh teams (freshness=1). Indeed, the overall disruption $D$, as well as old teams $D$, tend to \emph{decrease} with team size, supporting the interesting finding of Wu et al~\cite{large2019wu}. However, interestingly, we find that the disruption (originality) $D$ of the papers published by fresh teams tends to \emph{increase} with team size. The significance of this increasing trend is supported by the Kolmogorov-Smirnov test of the disruption distribution of different team sizes shown in Fig. S6 of SM. These results suggest that larger fresh teams play a more important role than small fresh teams in advancing science with new and original ideas and opportunities. Similar increasing trend can be observed when we examine the relation between multi-disciplinary impact and team size of fresh teams. Comparing the difference between fresh teams and old teams in $D$ and $M$, we also find that the advantage of fresh teams in creating original and diverse impact work is more prominent in larger teams.

In the above analysis, we define the team freshness as the fraction of new team members in a paper. It has been evaluated by calculating the fraction of nodes with no link to others in the collaboration network which represents the prior collaboration relations of the team members, see Fig. 1 (a) and (b). However, an alternative way to define freshness of a team could be by measuring the number of new collaboration relations (new links) created by the team. This can be regarded as a link freshness which could be of interest. This link freshness can be easily calculated by the fraction of missing links in the collaboration network which represents the prior collaboration relations of the team members (e.g, $2/6=1/3$ dashed links in Fig. 1(a)). To distinguish between these two types of freshness, we refer to them as node freshness $f_n$ (new collaborators) and link freshness $f_l$ (new collaborations) according to their calculations in the collaboration networks. An interesting question here is which types of freshness (node or link) is more important for the originality and impact diversity of the produced works. To test and answer this question, we show in Fig. 4a the scatter plot of link freshness versus node freshness, with circle size and color representing the mean disruption of the corresponding papers. Given a certain node freshness, higher link freshness is very little or even not associated with a higher disruption. This observation can be quantitatively supported using the Pearson correlation coefficient between link freshness and disruption for each given node freshness. One can see in the insets that the Pearson correlation between link freshness and disruption coefficients are very low and even in some cases negative, i.e., at the level of noise. For further support, see Fig. S7.

To better estimate the role of link freshness, we design a combined freshness measure $f_m$ as a weighted linear combination of node freshness and link freshness, with a tunable parameter controlling the relative weights of the two types of freshness (see Methods for more details). Next, we compute the Pearson correlation between the combined freshness $f_{m}$ and the disruption $D$. By tuning the relative weights of the two type of freshness, we find that the maximum correlation achieved with $f_{m}$, shown in Fig. 4b, is not significantly higher than the correlation between $f_n$ and $D$ (see significant test in SM Fig. S8). These results suggest that incorporating link freshness does not bring significant additional information for predicting originality. In Fig. 4c and 4d, we carried out similar analysis for the multi-disciplinary impact and find similar results.

We next consider another type of freshness of teams that we call here career freshness of team members. The career freshness of a team member can be measured by his/her career age, namely the number of years since he/she published the first paper. A shorter career age indicates a fresher scientist. The basic statistics of the mean career age of team members is shown in SM Fig. S9. We here further ask whether the career freshness of team members affects the originality and impact diversity of their produced works. In Fig. 5, we show the dependence of the mean disruption (originality) $D$ and multi-disciplinary impact $M$ on the mean career age of the team members of a paper. Surprisingly, we observe a decreasing trend in both cases. The decreasing trend is still present also when we fix the team freshness (as low freshness 0 and high freshness 1 in Fig. 5). Note that a similar trend is observed in Fig. S10 of SM when we use the mean productivity of team members to define the freshness of their careers. These results suggest that early career team members, i.e., higher career freshness, tend to produce more original and diverse impact research. There could be many reasons for this surprising phenomenon. One possible explanation for this is that researchers in earlier career stages are less likely to be trapped by concepts and general believes that are common in the scientific field, resulting in higher originality in their works.

In the literature, the tendency of individual teams in selecting new team members has been interestingly found to be related to the impact factor of the journals of the published studies~\cite{team2005guimera}. Thus, the question we ask here is how the team freshness is related to the citations impact of papers. To this end we analyze in Fig. S11 of SM how the team freshness affects the number of citations that a paper will receive. To be able to compare papers from different years, we calculate the citations received by a paper within 10 years of publication ($c_{10}$)~\cite{quantifying2016sinatra}. We show in Fig. S11 that papers produced by fresh teams tend to have smaller $c_{10}$ compared to old teams, which is consistent with the findings in ref.~\cite{team2005guimera}. It has been shown~\cite{reputation2014peterson} that the impact of a paper (number of citations) is positively correlated with the authors' cumulated reputation (measured by their productivity). To remove this effect, we consider only papers published by teams with similar team member productivity. After controlling this factor, we find that papers with different team freshness do not exhibit significant difference in $c_{10}$. Thus, our results suggest that the difference in the number of citations received by fresh teams and old teams can be attributed to team members' productivity instead of team freshness. Note that we have also shown in Fig. S4 that the increasing trend of disruption and multi-disciplinarity with team freshness is independent of team members' productivity.

\section{Discussion}
Despite intensive efforts in understanding team formation mechanism and the effect of team size on creativity, little is known about how the prior relations between team members affect the originality and impact diversity of the produced works. In this paper, we define the freshness of a team according to the fraction of team members without prior collaboration with other team members. We find that works of fresh teams are significantly more effective than works of old teams in disrupting science with original ideas and opportunities. In addition, the impact of the works produced by fresh teams is found to be more diverse, influencing multiple research areas. These two effects are found to be more prominent in larger teams. We also find that new team members is more powerful than new collaboration relations in predicting the originality and impact diversity of the resultant studies. Although the researchers in fresh teams have substantially smaller number of published papers than old teams, the team members' productivity is shown to be not a relevant factor that affects the increasing trend of originality and impact diversity with team freshness. We finally find that researchers in fresher careers have higher original and multidisciplinary papers. This could be since they are less likely to be trapped in conventional concepts and believes in the field, and tend to produce more original and multi-disciplinary research.

Our work supports the decrease of originality with team size discovered in ref.~\cite{large2019wu} and reveals a possible origin of this discovery. The decreasing trend of originality with team size could be explained due to having less freshness in larger teams (Fig. 1c-e). Indeed for fully fresh teams, both the originality and multidisciplinary impact increase significantly with team size (Fig. 3). In addition, we remark that the question of what comes first, the fresh team or the novel and original ideas, is not answered in our research. Our results discovered correlations between freshness and originality and between freshness and multi-disciplinary impact but not causality. It is possible that one (or more) author(s) had a novel idea or a novel problem and create a suitable new team to study it.

One of the main findings in this paper is that fresh teams play a critical role in creating original and diverse impact works. Funders and decision makers thus should encourage scientists forming fresh teams in research. Scientists themselves should also seize opportunities to interact with new colleagues for future collaboration as a new team. Based on the current work, several promising research extensions can be performed. A straightforward one is to investigate the performance of fresh teams in other activities such as research funding applications, software development, patent invention. Another interesting research direction is to study the mechanisms driving the formation of fresh teams. Finally, we remark that scientific collaboration is a complex phenomena, with the outcomes driven by multiple factors. Apart from the team freshness, the freshness of topic that the team studies is also a critical factor determining the quality of produced works~\cite{increase2019zeng,quantifying2017jia}. Therefore, identifying the inter-relations between team freshness and topic freshness would be an interesting topic for future study.

\section{Methods}
\small
\textbf{Data.} In this paper, we analyze the publication data of all journals of the American Physical Society (APS). The data contains 482,566 papers, ranging from year 1893 to year 2010. For the sake of author name disambiguation, we use the author name dataset provided by Sinatra et al. which is obtained with a comprehensive disambiguation process in the APS data~\cite{quantifying2016sinatra}. Eventually, a total number of 236,884 distinct authors have been matched. Another set of data that we analyzed in the supplementary materials is the computer science data obtained by extracting scientists' profiles from online Web databases~\cite{extraction2008tang}. The data contains 1,712,433 authors and 2,092,356 paper, ranging from year 1948 to year 2014. The author names in this data are already disambiguated. The third data set we analyzed is the publication data of Journal of the American Chemical Society (JACS). The data contains 59,913 papers, ranging from 1997 to 2017. We carry out the same name disambiguation process as in ref.~\cite{quantifying2016sinatra} and obtain 162,016 distinct authors. \\
\textbf{Disruption.} The disruption index is originally designed to identify destabilization and consolidation in patented inventions~\cite{dynamic2017funk}. In a recent article, it is extended to measure the originality of scientific papers~\cite{large2019wu}. The disruption, $D$, varies between $-1$ and $1$. $D=1$ of a paper indicates that all the paper's citing papers cite it but not any of its references. In this case, the paper is considered to disrupt science with new ideas and opportunities, corresponding to higher originality. If a paper has $D=-1$, all its citing papers not only cite it but also at least one of its references. In this case, the paper is devoted to further develop existing findings and ideas. The calculation of the disruption is illustrated in Fig. 1.\\
\textbf{Multi-disciplinary impact.} In this paper, we propose a simple index, called the multi-disciplinary impact $M$, to measure the diversity of the disciplines that a paper influences. Different from the various existing indexes that relies on external information such as disciplinary categories~\cite{general2007stirling,is2009porter}, our method is solely based on the citation relations. We define the multi-disciplinary impact of a paper as the probability of two successive citing papers from different disciplines. It can be easily obtained by calculating the fraction of temporal adjacent citing papers sharing no references apart from the focal paper. The multi-disciplinary impact, $M$, varies between $0$ and $1$, corresponding to narrow and diverse impact in disciplines, respectively. The calculation of the multi-disciplinary impact is illustrated in Fig. 1. Like the disruption index, the multi-disciplinary impact of a paper is not correlated with its citations (Fig. S2 in SM).\\
\textbf{Bootstrap disruption and bootstrap multi-disciplinary impact.} In the insets of Fig. 2, we compare the distributions of bootstrap disruption of papers with team freshness $0$ and $1$. The bootstrap disruption is obtained by random sampling of papers' disruption such that each paper's disruption has an equal chance to be selected and can be selected over and over again. The distributions are obtained by performing 1000 realizations of bootstrap disruption. The bootstrap multi-disciplinary impact in the insets of Fig. 2 is obtained similarly.\\
\textbf{Combined freshness measure.} The node freshness of a team $f_n$ is defined as the fraction of nodes with no link to others in the collaboration network which represents the prior collaboration relations of the team members. The link freshness of a team $f_l$ is defined as the fraction of missing links in the collaboration network representing the prior collaboration relations of the team members. Denoting the combined freshness measure as $f_m$, it is computed as $f_{m}=\lambda f_{n}+(1-\lambda)f_{l}$ where $\lambda$ is a tunable parameter between 0 and 1. In Fig. 4bd, we show the maximal Pearson correlations that can be achieved by adjusting $\lambda$.

\clearpage

\noindent  \textbf{Acknowledgments} \\
This work is supported by the National Natural Science Foundation of China under Grant (Nos. 71843005 and 71731002). SH thanks the Israel Science Foundation and the NSF-BSF for financial support.

\noindent \\ \textbf{Author contributions}\\
AZ, SH designed the research, AZ performed the experiments, YF, ZD, YW contributed analytic tools, AZ, SH analyzed the data, all authors wrote the manuscript.

\noindent \\ \textbf{Competing financial interests.} The authors declare no competing financial interests.\\

\noindent \textbf{Data and materials availability.} The APS data can be downloaded via \url{https://journals.aps.org/datasets} and the computer science data can be downloaded via \url{https://www.aminer.cn/aminernetwork}. Other related, relevant data are available from the corresponding author upon reasonable request.

\clearpage
\section*{Figures}
\begin{center}
  \centering
  \includegraphics[width=16cm]{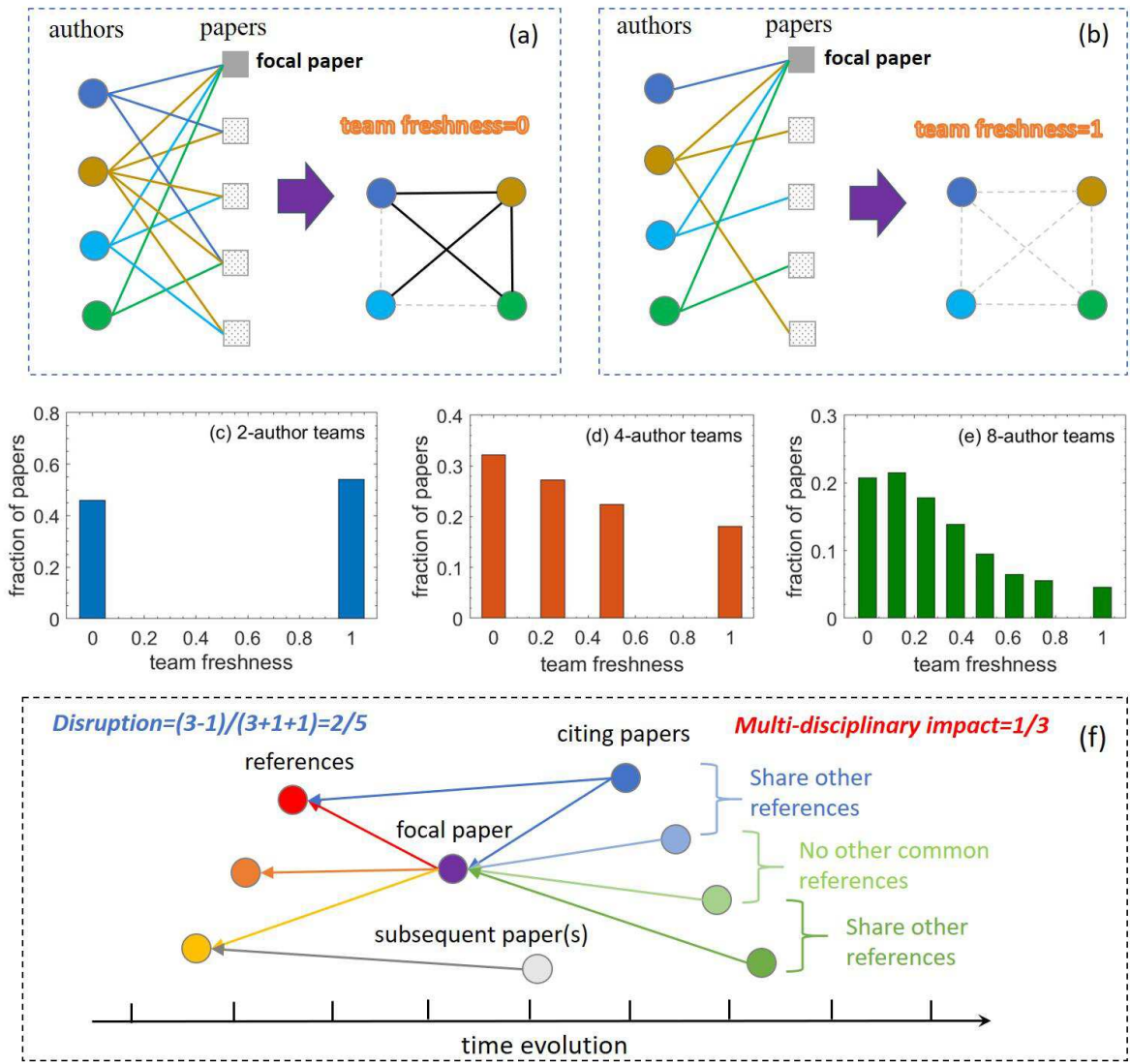}\\
  	\bigskip
	\setbox0\vbox{\makeatletter
		\let\caption@rule\relax
\captionof{figure}[short caption]{\textbf{Illustration of freshness of teams, disruption of papers and multi-disciplinary impact of papers.} (a) and (b) The four authors (circles) in the toy bipartite network are the authors of the focal paper (filled square). The other papers (empty squares) are the papers published by the four authors before the focal paper. The collaboration network of these authors before the focal paper can be constructed, with solid and dashed lines representing the existing and missing links respectively. The fraction of nodes with zero collaboration links in the prior network is defined as the freshness of the team in the focal paper. Accordingly, the team of the focal paper in (a) has freshness 0 and the team of the focal paper in (b) has freshness 1. In our analysis, we study all the 482,566 papers published during the years 1893-2010 by the American Physical Society (APS). (c)-(e) show freshness distribution for 2-author papers, 4-author papers and 8-author papers, respectively. One immediate observation is that completely fresh teams are less common in larger teams. (f) Demonstration of calculating disruption~\cite{dynamic2017funk,large2019wu} and multi-disciplinary impact in a citation network. The citation network consists of a focal paper, its references (outgoing links) and its citing papers (incoming links). The disruption aims to measure the originality of a paper. To calculate the disruption of the focal paper, one should first calculate the difference between the number of its citing papers that do not cite its references and the number of its citing papers that cite its references. The disruption is obtained by dividing this difference by the number of all citing papers plus the number of subsequent papers of the focal paper that do not cite it but its references. In the example, disruption of the focal paper is (3-1)/(3+1+1)=2/5. The disruption varies between $-1$ and $1$, with larger disruption corresponding to higher originality. The multi-disciplinary impact aims to measure the diversity of the areas that a paper influences. We define it here as the fraction of temporal adjacent citing papers sharing no references apart from the focal paper (see Methods for more details). In the example, the focal paper has 4 citing papers, resulting in 3 adjacent pairs in time. Among these 3 pairs, 1 pair share no other common references apart from the focal paper and 2 pairs share other references apart from the focal paper. Thus, the multi-disciplinary impact of the focal paper is 1/3. This index varies between $0$ and $1$, corresponding to narrow and diverse impact in disciplines, respectively.}
\global\skip1\lastskip\unskip
		\global\setbox1\lastbox
	}
	\unvbox0
	\setbox0\hbox{\unhbox1\unskip\unskip\unpenalty
		\global\setbox1\lastbox}
	\unvbox1
	\vskip\skip1
\end{center}\label{fig1}

\clearpage
\begin{figure}[h!]
  \centering
  \includegraphics[width=16.5cm]{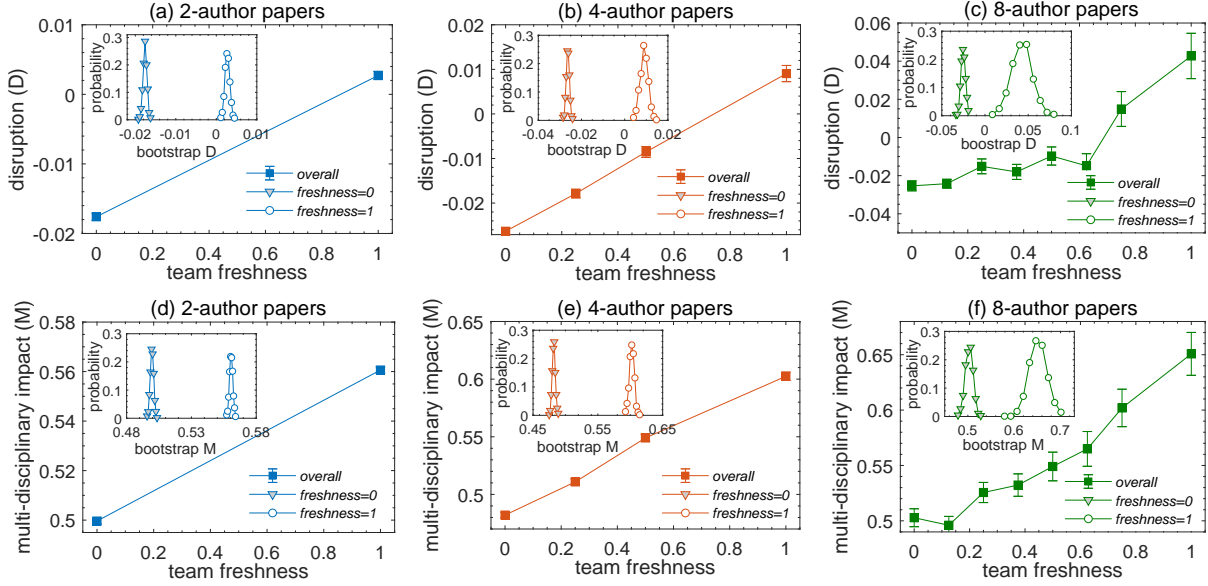}\\
  \caption{\textbf{Fresh teams create more original and multi-disciplinary research.} Shown are the dependence of the disruption (originality)
$D$ and multi-disciplinarity $M$ of papers on the team freshness, for (a)(d)
2-author papers, (b)(e) 4-author papers, (c)(f) 8-author papers, respectively. The results suggest that both originality and multi-disciplinarity significantly increase with team freshness. The insets show the distributions of 1000 realizations of bootstrap disruption or bootstrap multi-disciplinarity. A remarkable difference, i.e., high significance, can be observed between the distributions of $D$ of papers with team freshness 0 and 1. The $p$-values of the Kolmogorov-Smirnov test of difference (in $D$ or $M$) between papers with team freshness 0 and 1 are all smaller than 0.001 (For other freshness values see Fig. S1).}\label{fig2}
\end{figure}

\clearpage
\begin{figure}[h!]
  \centering
  \includegraphics[width=14cm]{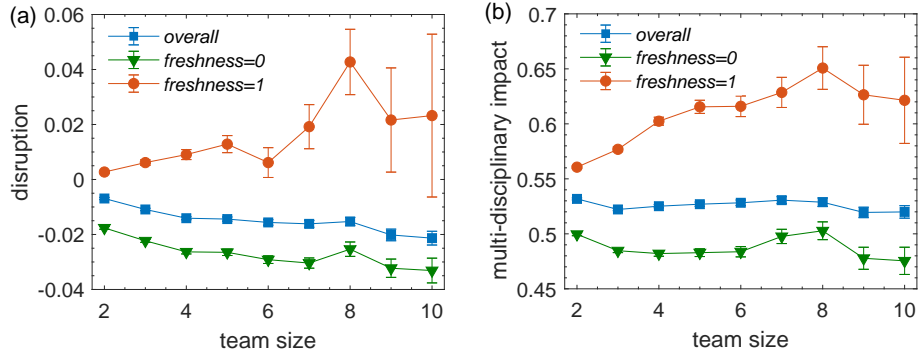}\\
  \caption{\textbf{Difference between fresh and old teams is amplified in larger teams.} (a) Plot of the mean disruption (originality) $D$ of papers of different team sizes (overall), showing decreasing trend as discovered in ref.~\cite{large2019wu}. For each team size, we also study the mean disruption $D$ of papers published by old teams (freshness=0) and fresh teams (freshness=1). Surprisingly, in contrast to the overall papers, for papers of freshness $1$, $D$ \emph{increases} with team size. This suggests that the overall decreasing disruption is due to the dominant non-fresh teams. (b) Plot of the mean multi-disciplinary impact $M$ of papers of different team sizes (overall). For each team size, we also study the mean multi-disciplinary impact $M$ of papers published by old teams (freshness=0) and fresh teams (freshness=1). Similar to the results of $D$, one can see that the difference in $M$ between fresh and old teams is amplified in larger teams.}\label{fig3}
\end{figure}

\clearpage
\begin{figure}[h!]
  \centering
  \includegraphics[width=14cm]{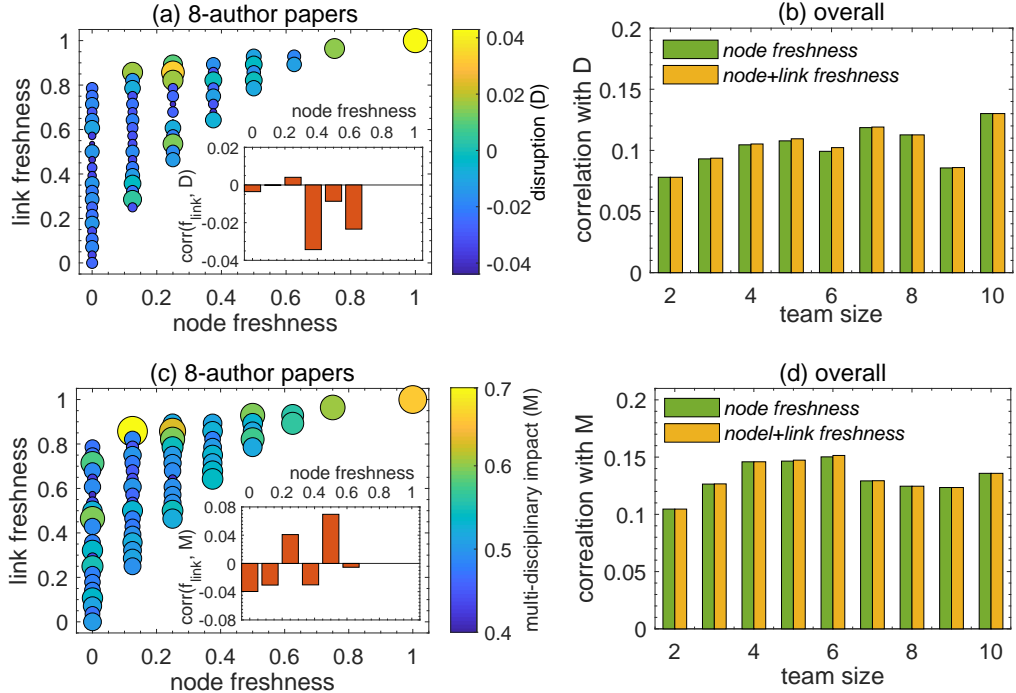}\\
  \caption{\textbf{Team freshness defined by new team members and new collaboration relations.} We define the node freshness of a paper as the fraction of new team members, and the link freshness as the fraction of new collaboration relations in the team. (a) The scatter plot of link freshness versus node freshness for 8-author papers, with the circle size and the color representing the mean originality (disruption) of the corresponding papers (Similar results for other team sizes are given in Fig. S7). Given a certain node freshness of a paper, it is seen that higher link freshness is not associated with a higher disruption. This finding is supported by directly calculating the Pearson correlation between link freshness and originality (disruption) for each node freshness (as shown in inset). (b) The Pearson correlation of node freshness and originality (disruption) for papers of different team sizes. For comparison, we calculate the maximum Pearson correlation when we consider team freshness as a weighted linear combination of node and link freshness. The results suggest that incorporating link freshness does not bring significant additional information for predicting disruption. (c) The scatter plot of link freshness versus node freshness for 8-author papers, with the circle size and the color representing the mean multi-disciplinary impact (Similar results for other team sizes are given in Fig. S7). The Pearson correlation between link freshness and multi-disciplinarity for each node freshness is shown in inset. (d) The Pearson correlation of node freshness and multi-disciplinary impact for papers of different team sizes. We show also the maximum Pearson correlation of the weighted linear combination of both node and link freshness.}\label{fig4}
\end{figure}

\clearpage
\begin{figure}[h!]
  \centering
  \includegraphics[width=16.5cm]{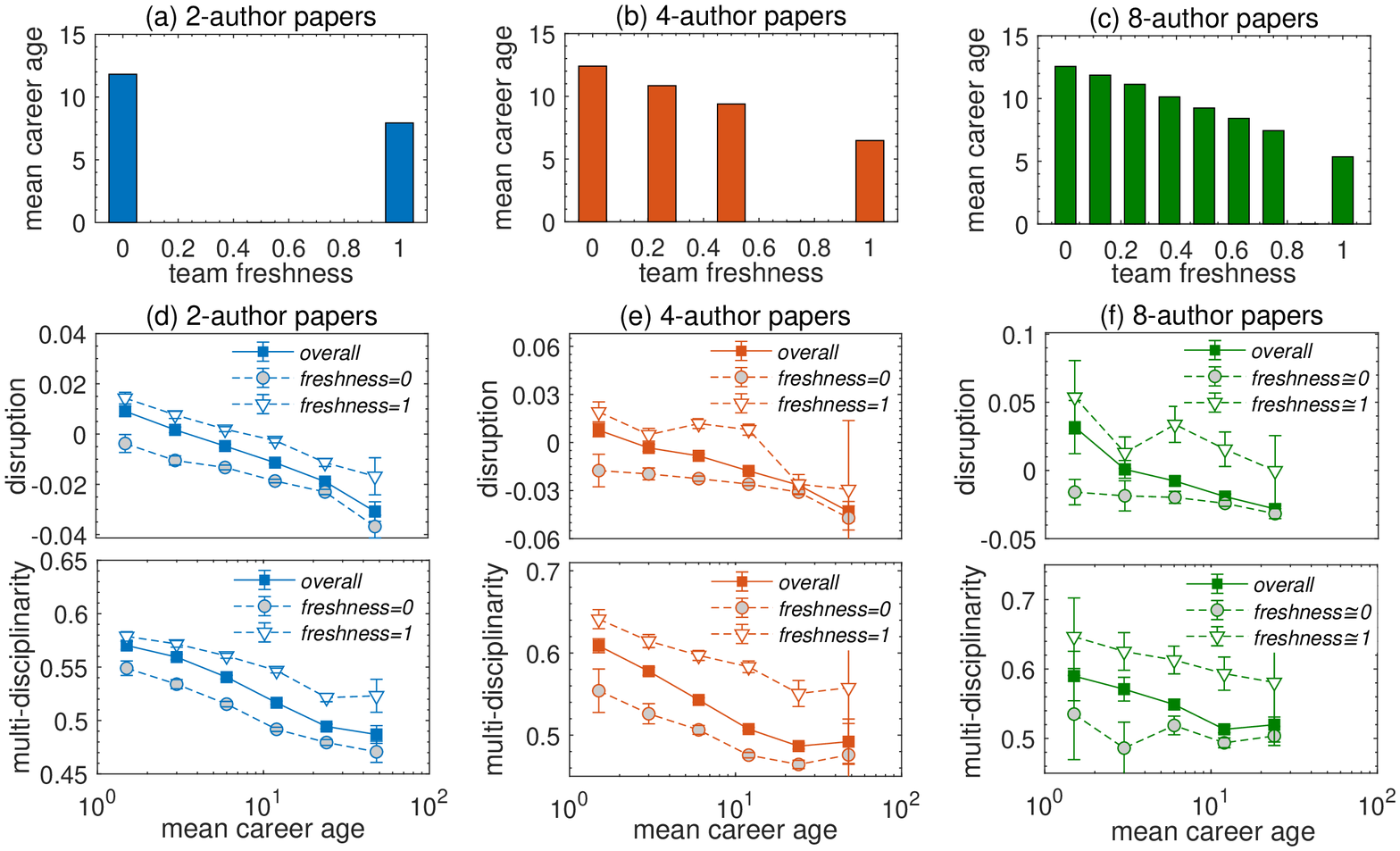}\\
  \caption{\textbf{Freshness of team member's careers.} In the top three panels, we show the mean career age of team members in team sizes of (a) 2, (b) 4 and (c) 8 members. The career age of a team member is defined as the number of years after he/she publishes the first paper. The results suggest that scientists in fresh teams tend to have smaller career age than those in old teams. In the second and third row, we show respectively the dependence of the mean disruption (originality) $D$ and multi-disciplinarity $M$ on team members' mean career age, in (d) 2-author papers, (e) 4-author papers and (f) 8-author papers. To remove the effect of team freshness, we show also that the curves for old teams (freshness=0) and fresh team (freshness=1) behave similarly, decrease with mean career age. For better statistics in 8-author papers, we take freshness$\leq0.25$ as freshness$\cong0$ and freshness$\geq0.85$ as freshness$\cong1$. The results suggest that younger scientists tend to produce more original and multi-disciplinary works. }\label{fig5}
\end{figure}

\clearpage
\begin{center}
{\large\bfseries Supplementary Information}\\[8pt]
{\large The critical role of fresh teams in creating original and multi-disciplinary research}\\[8pt]
\small An Zeng, Ying Fan, Zengru Di, Yougui Wang, Shlomo Havlin\\
\end{center}


\begin{figure}[h!]
  \centering
  \includegraphics[width=16.5cm]{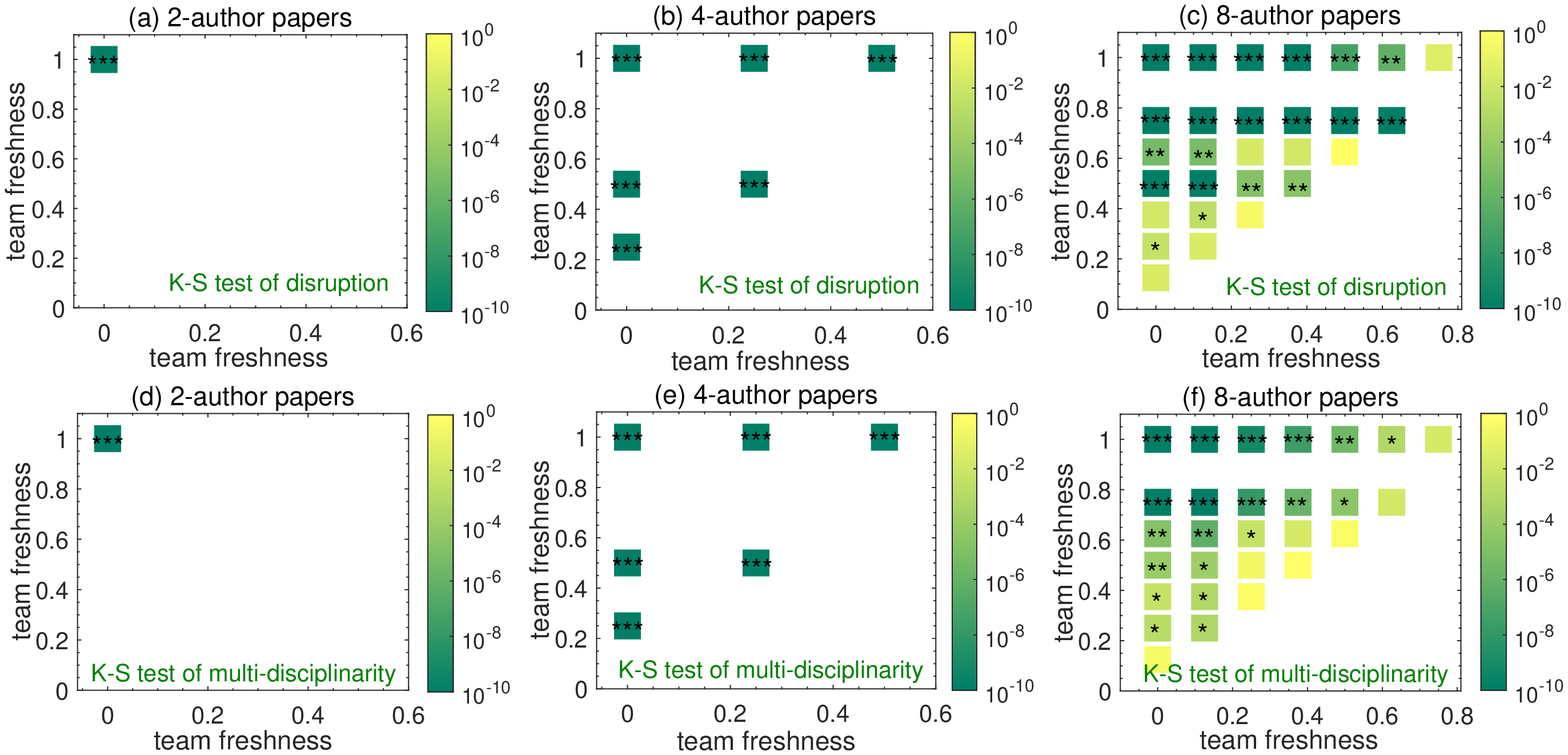}\\
  \textbf{Figure S1. Kolmogorov-Smirnov test of the significance of differences (in disruption $D$ or multi-disciplinarity $M$) between papers with different team freshness.} (a)-(c) We test the differences between the distributions of disruption for each pair of team freshness from zero to one using the Kolmogorov-Smirnov test. The results for (a) 2-author papers, (b) 4-author papers, and (c) 8-author papers are shown. (d)-(f) We also test differences between distributions of multi-disciplinarity for each pair of team freshness from zero to one using the Kolmogorov-Smirnov test. The results for (d) 2-author papers, (e) 4-author papers, and (f) 8-author papers are shown. For more strict test of the distribution differences, we take the actual values of disruption and multi-disciplinarity instead of averaged values from bootstrap for the Kolmogorov-Smirnov test. Asterisks under the numbers indicate $P$ values. $*P\leq0.1$, $**P\leq0.01$, $***P\leq0.001$. Almost all pairs of tested distributions significantly differ from one another.\label{FigS}
\end{figure}

\clearpage
\begin{figure}[h!]
  \centering
  \includegraphics[width=16.5cm]{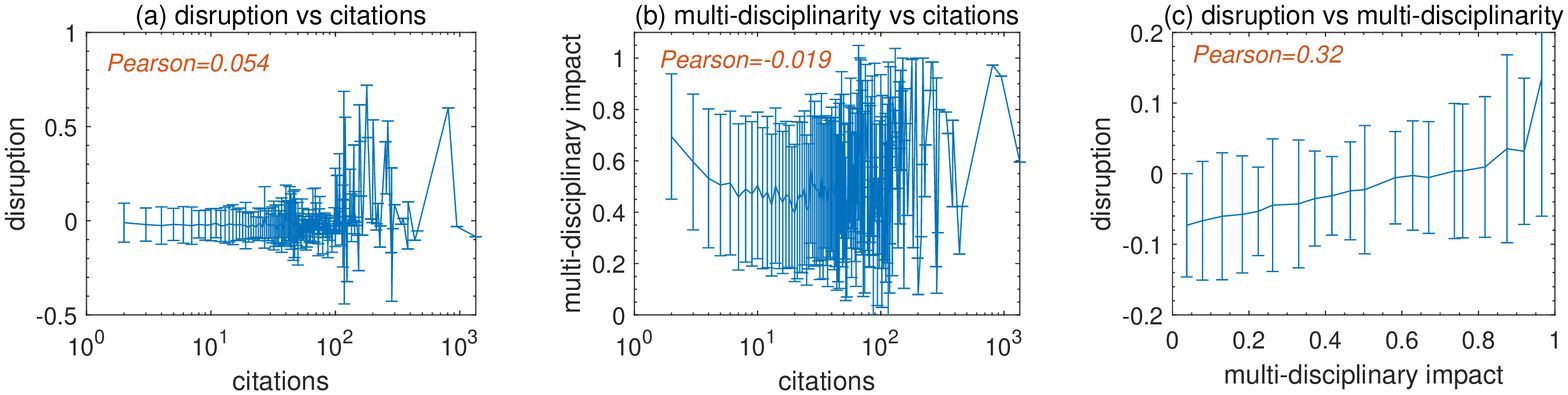}\\
  \textbf{Figure S2. Correlation between disruption and multi-disciplinary impact.} (a) The average disruption of papers with different citations. No significant trend can be observed, indicating that disruption (representing originality) is not correlated with citations. This finding is supported by directly calculating the Pearson correlation coefficient between disruption and citations, which results in 0.054 correlation coefficient. (b) The average multi-disciplinary impact for papers with different citations. Similar to disruption, no significant correlation is observed. This is confirmed by the Pearson correlation coefficient -0.019. (c) The average disruption for papers of different multi-disciplinary impact. An increasing trend is observed, indicating that papers with higher multi-disciplinary impact generally have higher disruption. However, the Pearson correlation coefficient is only 0.32, suggesting that these two measures are still largely different from each other.\label{FigS}
\end{figure}

\clearpage
\begin{figure}[h!]
  \centering
  \includegraphics[width=16.5cm]{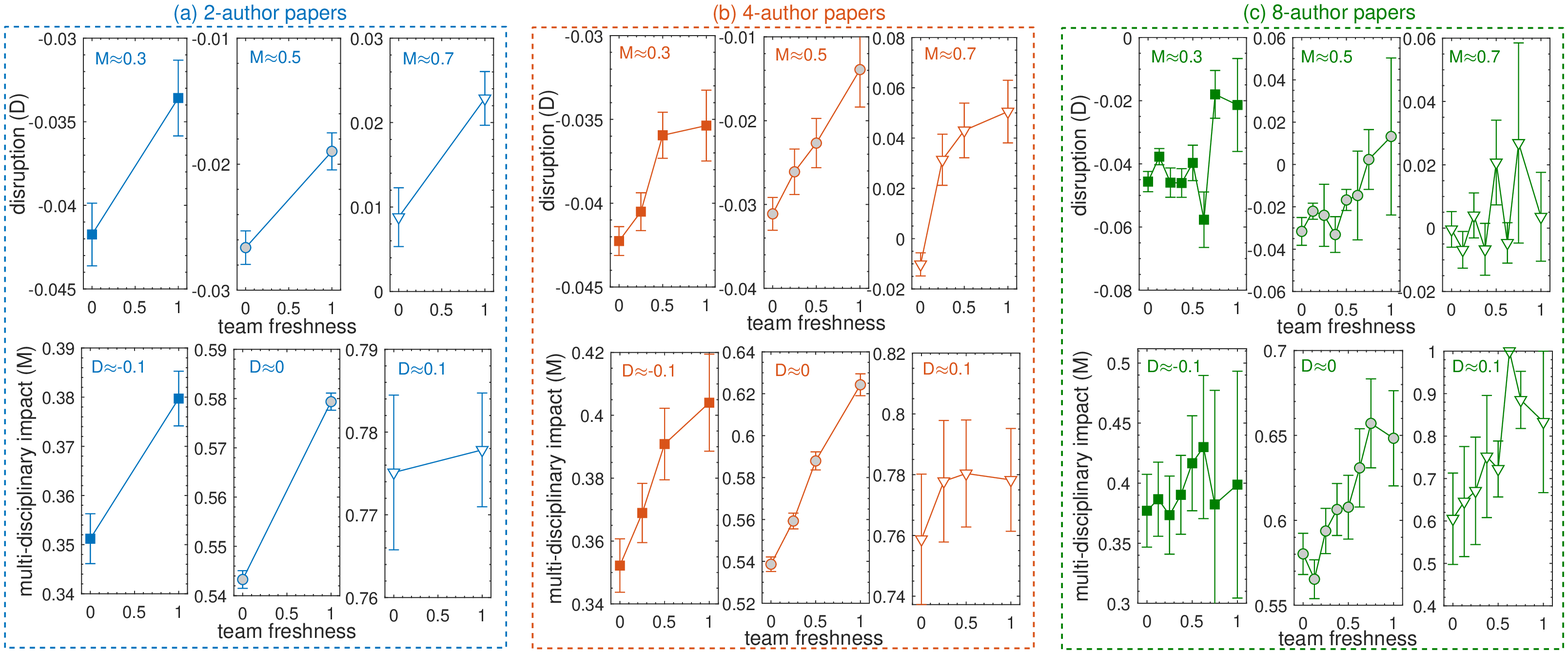}\\
  \textbf{Figure S3. Disruption versus team freshness when controlling multi-disciplinarity, and multi-disciplinarity versus team freshness when controlling disruption.} As disruption $D$ is to some degree correlated with multi-disciplinary impact $M$, a crucial question is whether the increasing trends of both indexes with team freshness measure the same feature. To address this question, we study the dependence of disruption on team freshness when controlling multi-disciplinarity, $M$. Specifically, we fix (top figures) multi-disciplinarity $M$ as $0.3\pm0.01$, $0.5\pm0.01$ and $0.7\pm0.01$ in (a) 2-author papers, (b) 4-author papers, and (c) 8-author papers, respectively. The results suggest that the increasing trend of disruption with team freshness occurs even for fixed multi-disciplinarity. In addition, we study (bottom figures) the dependence of multi-disciplinarity on team freshness while controlling disruption. We fix disruption $D$ as $-0.1\pm0.01$, $0\pm0.01$ and $0.1\pm0.01$ in (a) 2-author papers, (b) 4-author papers, and (c) 8-author papers, respectively. The results suggest that the increasing trend of multi-disciplinarity with team freshness occurs even for fixed disruption.\label{FigS}
\end{figure}

\clearpage
\begin{figure}[h!]
  \centering
  \includegraphics[width=16.5cm]{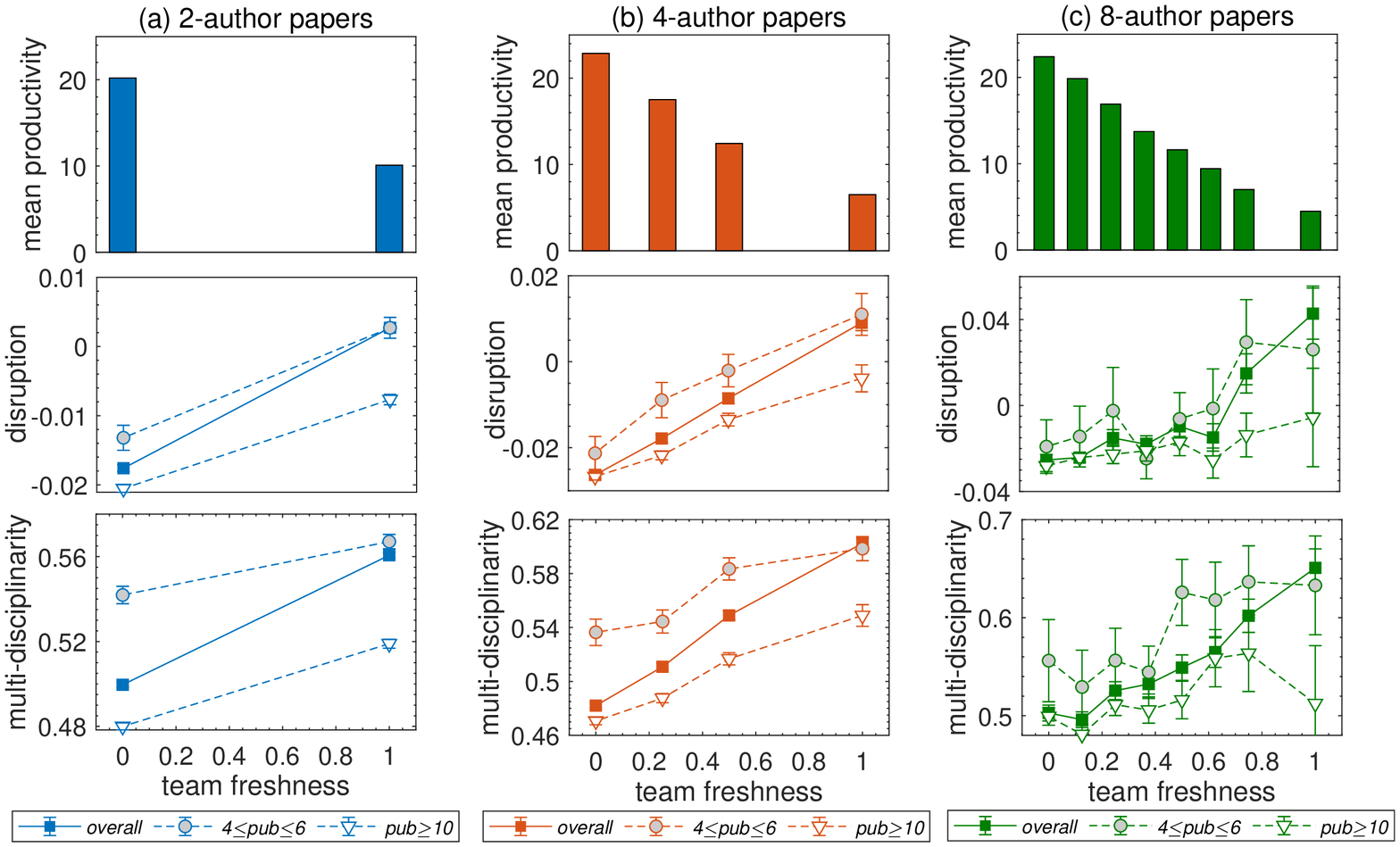}\\
  \textbf{Figure S4. Freshness in teams with similar team member productivity.} In the top three panels, we show the mean prior productivity of team members in teams with size (a) 2, (b) 4 and (c) 8, respectively. The prior productivity of a team member is defined as the number of his/her papers before he/she publishes the focal paper. The results suggest that scientists in fresh teams tend to have lower prior productivity than those in old teams. In the second row, we show the increase of originality (disruption) with freshness for papers of team size 2, 4 and 8. To support the trend, we also plot the results when controlling the team member productivity. Accordingly, we add results of teams with mean team member productivity between 4 and 6, and teams with mean team member productivity larger than 10. In both cases the increased trend of disruption with freshness are still present. In the bottom three panels, we show the increase of multi-disciplinary impact with freshness in papers of team size 2, 4 and 8. Again, these results support the trend when controlling team member productivity.\label{fig5}
\end{figure}

\clearpage
\begin{figure}[h!]
  \centering
  \includegraphics[width=16.5cm]{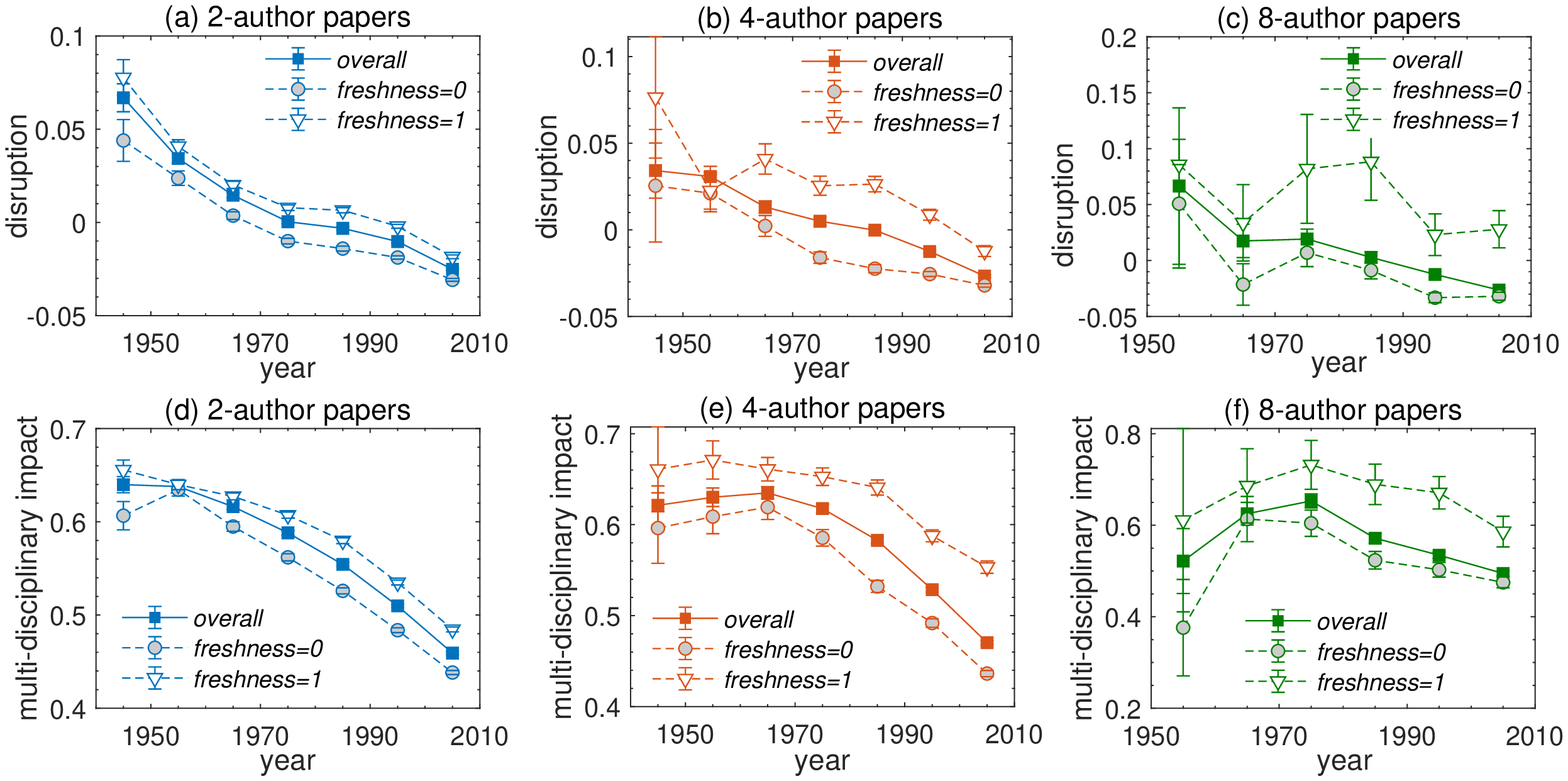}\\
  \textbf{Figure S5. Time evolution of originality and multi-disciplinary impact of papers.} The time evolution of the mean disruption (originality) of (a) 2-author papers, (b) 4-author papers, and (c) 8-author papers. The time evolution of the multi-disciplinary impact of (d) 2-author papers, (e) 4-author papers, and (f) 8-author papers. The results suggest that both indexes decrease (at least in the last 50 years) with time, which is also observed in ref.~\cite{large2019wu}. For comparison, also the papers published by old teams (freshness=0) and fresh teams (freshness=1) are shown in each panel, showing a similar pattern. It is also seen that fresh teams systematically have higher originality and multi-disciplinarity than old teams.\label{fig6}
\end{figure}

\clearpage
\begin{figure}[h!]
  \centering
  \includegraphics[width=16.5cm]{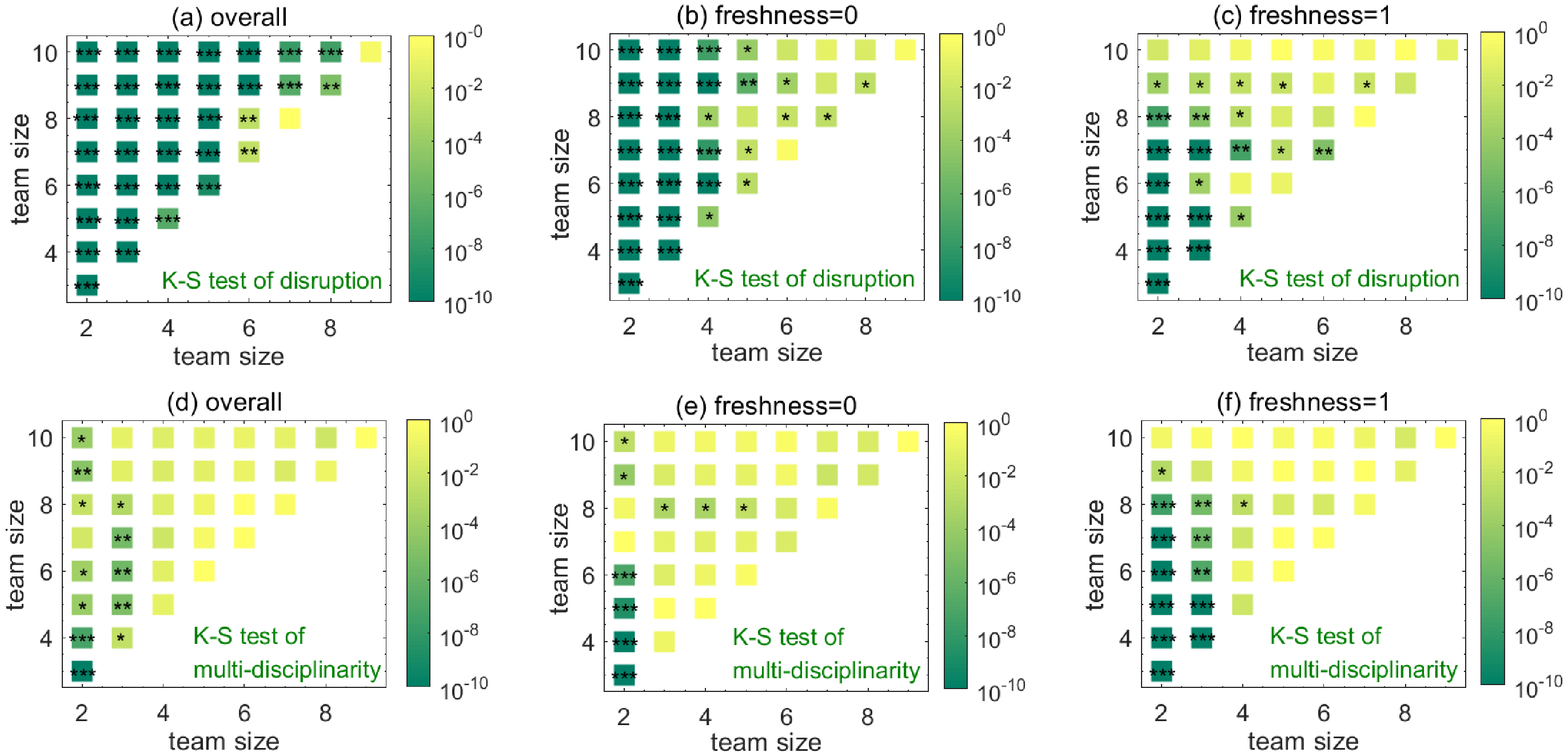}\\
  \textbf{Figure S6. Kolmogorov-Smirnov siginifcance test of difference of disruption $D$ or multi-disciplinarity $M$ between papers of different team sizes.} We test the differences between distributions of disruption for each pair of team sizes from two to ten using the Kolmogorov-Smirnov test. We also test differences between distributions of multi-disciplinarity for each pair of team size. In addition to the overall results in (a) and (d), we show also the results for papers with freshness=0 in (b)(e) and papers with freshness=1 in (c)(f). For more strict test of the distribution differences, we take the actual values of disruption and multi-disciplinarity instead of averaged values from bootstrap for the Kolmogorov-Smirnov test. Asterisks under the numbers indicate $P$ values: $*$ for $P\leq0.1$, $**$ for $P\leq0.01$, $***$ for $P\leq0.001$. Almost all pairs of tested distributions of disruption significantly differ from one another, indicating that team size is a critical factor affecting disruption (originality). However, most of the tested distributions of multi-disciplinarity are not significantly different from each other, except when comparing team size 2 and 3 with other team sizes.\label{FigS}
\end{figure}

\clearpage
\begin{figure}[h!]
  \centering
  \includegraphics[width=16.5cm]{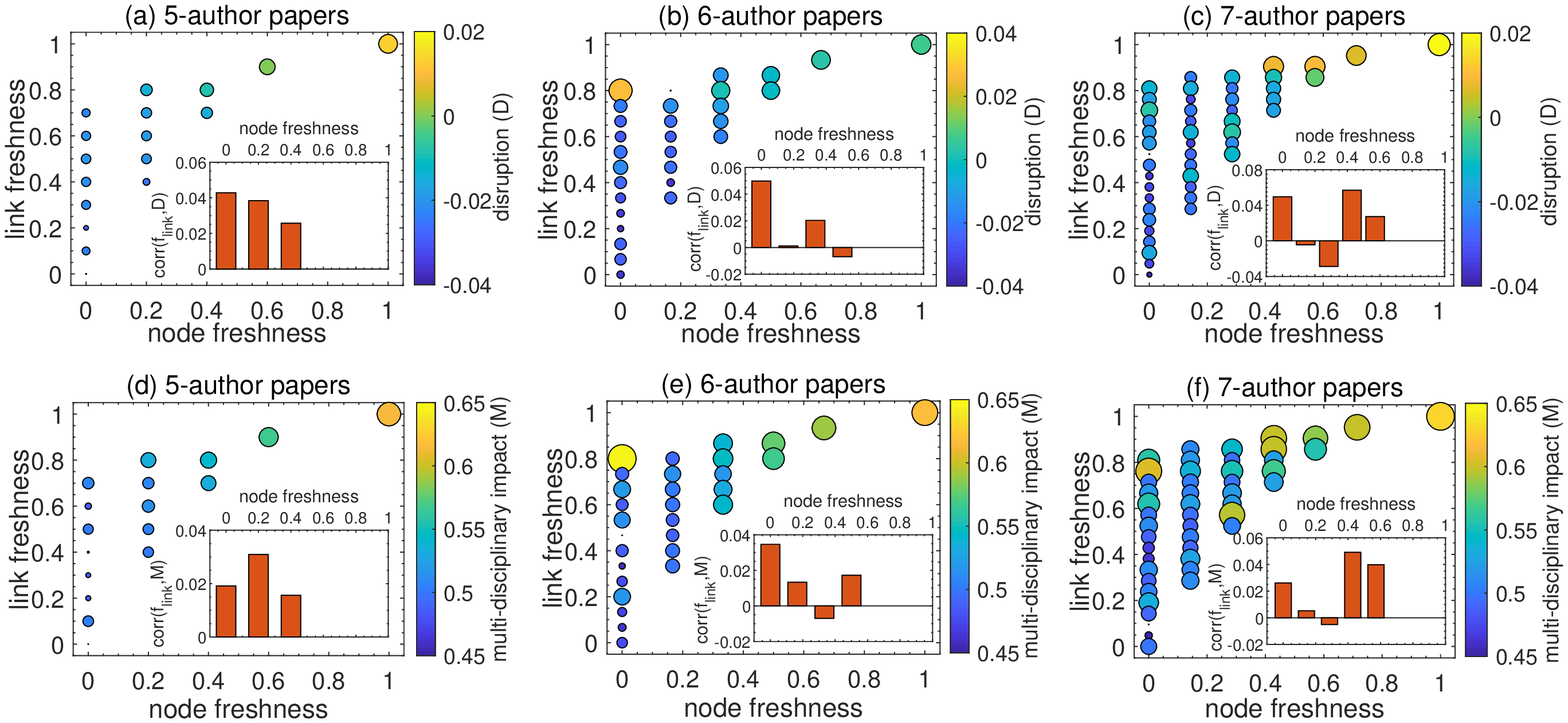}\\
  \textbf{Figure S7. Node freshness versus link freshness for papers of 5, 6 or 7 authors.} In the manuscript, we show the dependence of disruption and multi-disciplinary impact of 8-author papers on node freshness and link freshness. In the APS data sets, the number of 8-author papers is relatively small, 6965 in total. For better statistics, we examine here also papers of 5, 6, 7 authors (There are 27895 papers of 5 authors, 16858 papers of 6 authors, 10487 papers of 7 authors). (a)-(c) The scatter plot of link freshness versus node freshness for (a) 5-author, (b) 6-author, (c) 7-author papers, with the circle size and the color representing the mean originality (disruption) of the corresponding papers. The Pearson correlation between link freshness and disruption for each node freshness is very small and shown in the insets. (d)-(f) The scatter plot of link freshness versus node freshness for (d) 5-author, (e) 6-author, (f) 7-author papers, with the circle size and the color representing the mean multi-disciplinary impact of the corresponding papers. The Pearson correlation between link freshness and multi-disciplinarity for each node freshness is very small and shown in the insets.\label{fig4}
\end{figure}

\clearpage
\begin{figure}[h!]
  \centering
  \includegraphics[width=14cm]{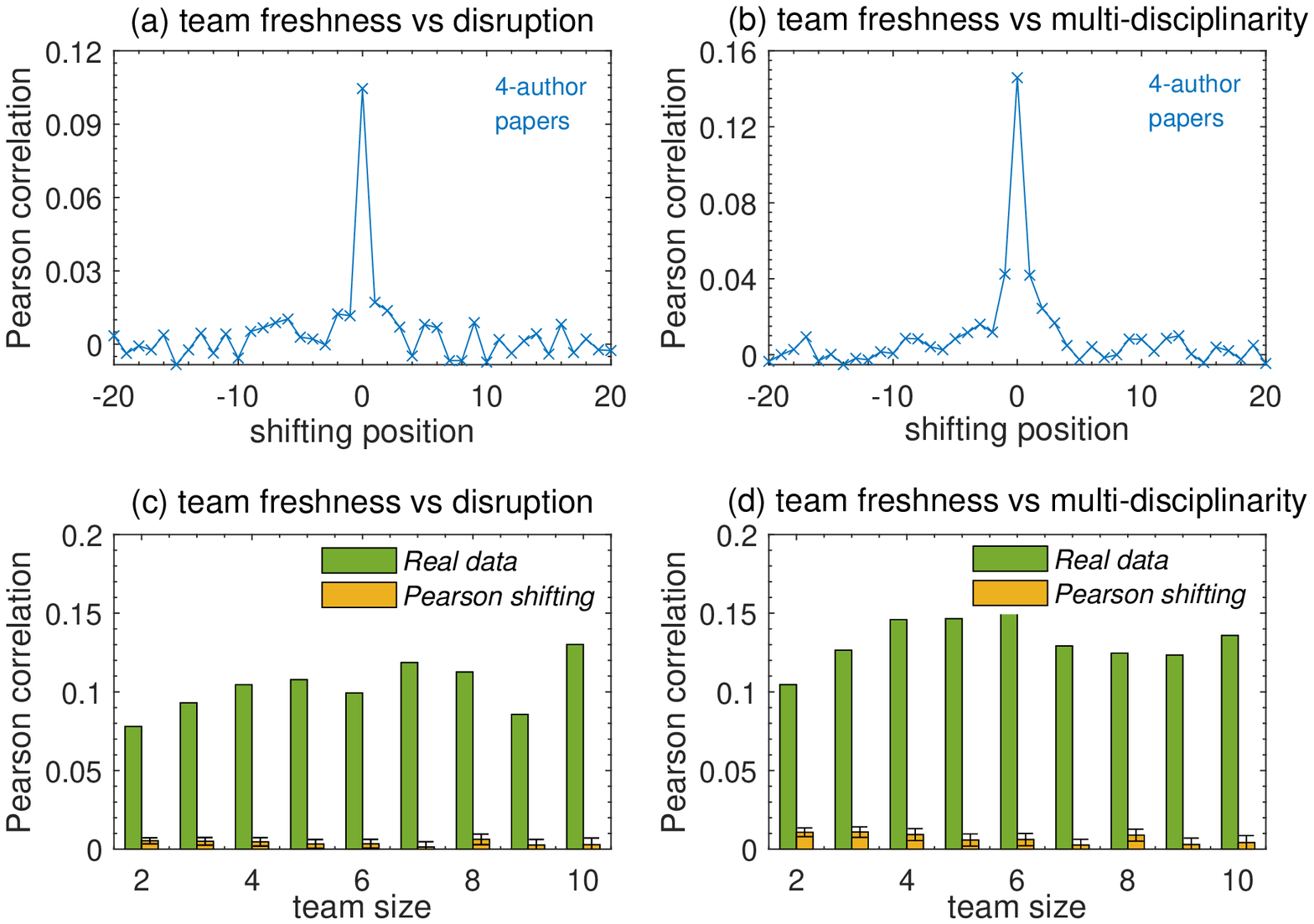}\\
  \textbf{Figure S8. Significance of the correlation of team freshness with disruption and multi-disciplinarity.} (a) To support the significance of the correlation between team freshness and originality (disruption), see Fig. 4b, we conduct the Pearson shifting test in which the Pearson correlation is calculated after shifting all elements in the vector of team freshness by certain number of positions. We take all 4-author papers and calculate the Pearson correlation (between the vectors of freshness and originality) with shifting one vector with respect to the other by 20 positions (-20 to 20). The true correlation is without movements while each movement is like a random correlations, showing therefore the level of noise. The sharp peak at shifting zero suggests that the correlation of team freshness and originality (disruption) in the original data is indeed significant. (b) The shifting Pearson correlation between team freshness and multi-disciplinarity. A sharp peak at shifting zero can be observed. (c) The Pearson correlation between team freshness and originality (disruption) for papers with different team sizes. To support the significance of the correlation, we show the averaged Pearson correlation over different number of shifting positions from -20 to 20 (without shifting 0). The results show that the Pearson correlation after shifting becomes much lower, indicating the significance of the Pearson correlation in real data. (d) The Pearson correlation between team freshness and multi-disciplinarity for papers with different team sizes. The Pearson shifting test also indicates the significance of the Pearson correlation in real data.\label{fig4}
\end{figure}

\clearpage
\begin{figure}[h!]
  \centering
  \includegraphics[width=16.5cm]{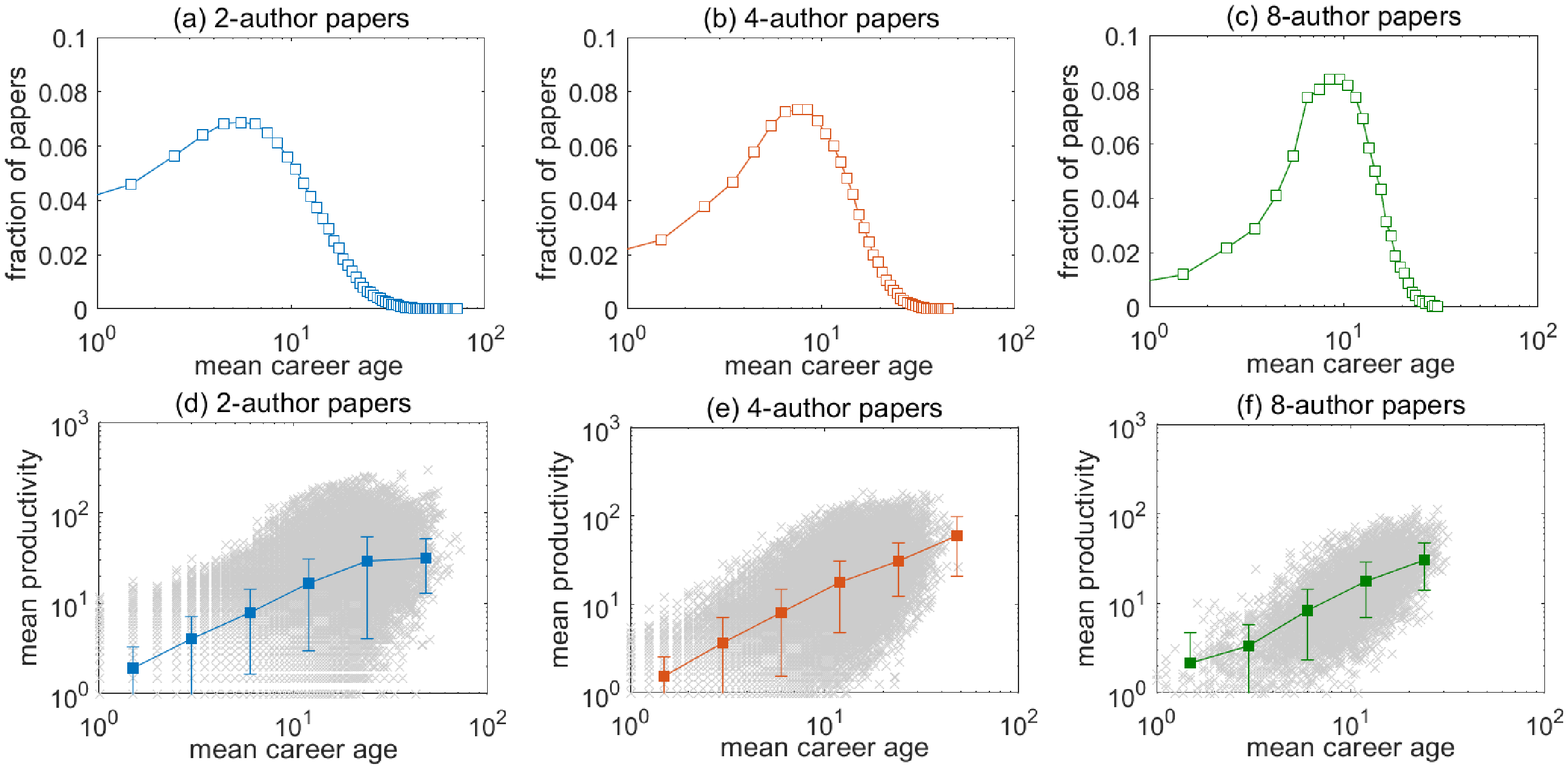}\\
  \textbf{Figure S9. Basic statistics of mean career age of team members.} The mean career age of team members is the average career age of the team members. The mean productivity of team members is the average number of papers published by team members before the focal paper. (a)-(c) Distributions of mean career age of (a) 2-author papers, (b) 4-author papers, and (c) 8-author papers. A peak can be observed for each distribution (peaking at age=5.5 years for 2-author papers, peaking at age=8 years for 4-author papers, peaking at age=9 years for 8-author papers). (d)-(f) Scatter plot of the mean productivity versus mean career age for (d) 2-author papers, (e) 4-author papers, and (f) 8-author papers. The averaged curves indicate a clear positive correlation between mean productivity and mean career age of team members.\label{fig5}
\end{figure}

\clearpage
\begin{figure}[h!]
  \centering
  \includegraphics[width=16.5cm]{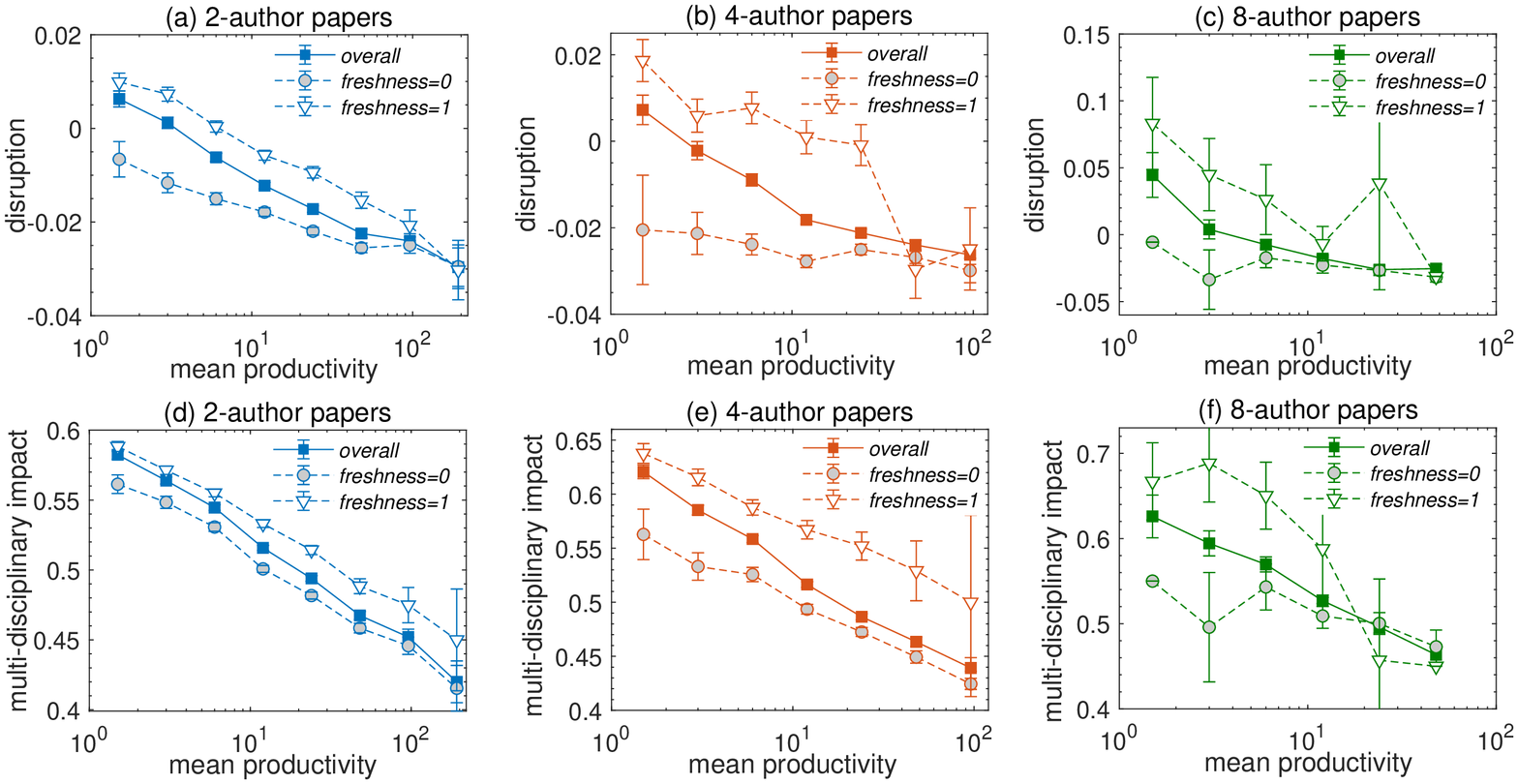}\\
  \textbf{Figure S10. Freshness based on productivity of team member's careers.} The mean prior productivity of team members also represents the freshness of team members' careers. A small mean productivity indicates that team members are in the early stage of their careers, namely a fresh career. We show the dependence of the mean disruption (originality) $D$ on team member productivity, in (a) 2-author papers, (b) 4-author papers and (c) 8-author papers. To remove the effect of team freshness, we show the dependence of disruption on team member productivity for old teams (freshness=0) and fresh team (freshness=1), respectively. The results suggest that fresher career team members tend to produce more original works. We show also the dependence of the multi-disciplinary impact $M$ on team member productivity, in (d) 2-author papers, (e) 4-author papers and (f) 8-author papers. The results of old teams (freshness=0) and fresh team (freshness=1) are also shown. The results suggest that fresher career teams in fresher careers tend to produce works with more diverse impact.\label{fig6}
\end{figure}

\clearpage
\begin{figure}[h!]
  \centering
  \includegraphics[width=16.5cm]{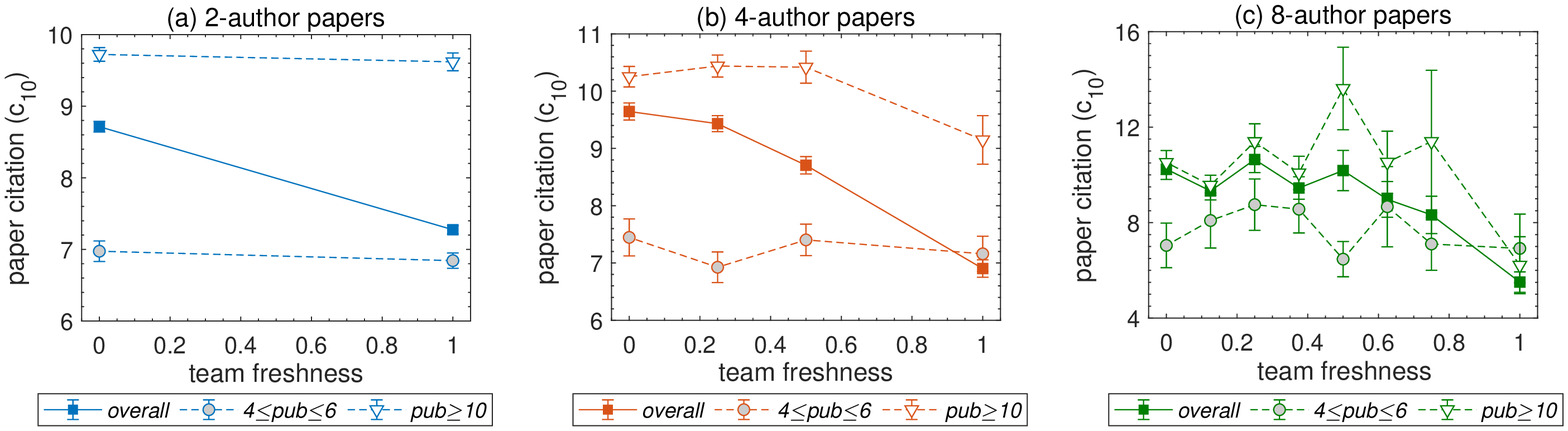}\\
  \textbf{Figure S11. Relation between team freshness and paper citations.} For each paper, we calculate the number of citations it received during ten years after its publication, denoted as $c_{10}$. In this figure, we show the citation $c_{10}$ for papers of different team freshness, in (a) 2-author papers, (b) 4-author papers, (c) 8-author papers, respectively. It is seen that citation decreases with freshness, which is consistent with the findings in ref.~\cite{team2005guimera}. For comparison, we consider papers published by the teams with team members' prior productivity around $5$ ($4\leq pub\leq6$) and higher than $10$ ($pub\geq10$). After controlling team members' productivity, one can see, however, that papers with different team freshness does not exhibit significant difference in $c_{10}$.\label{fig6}
\end{figure}

\clearpage
\begin{figure}[h!]
  \centering
  \includegraphics[width=16.5cm]{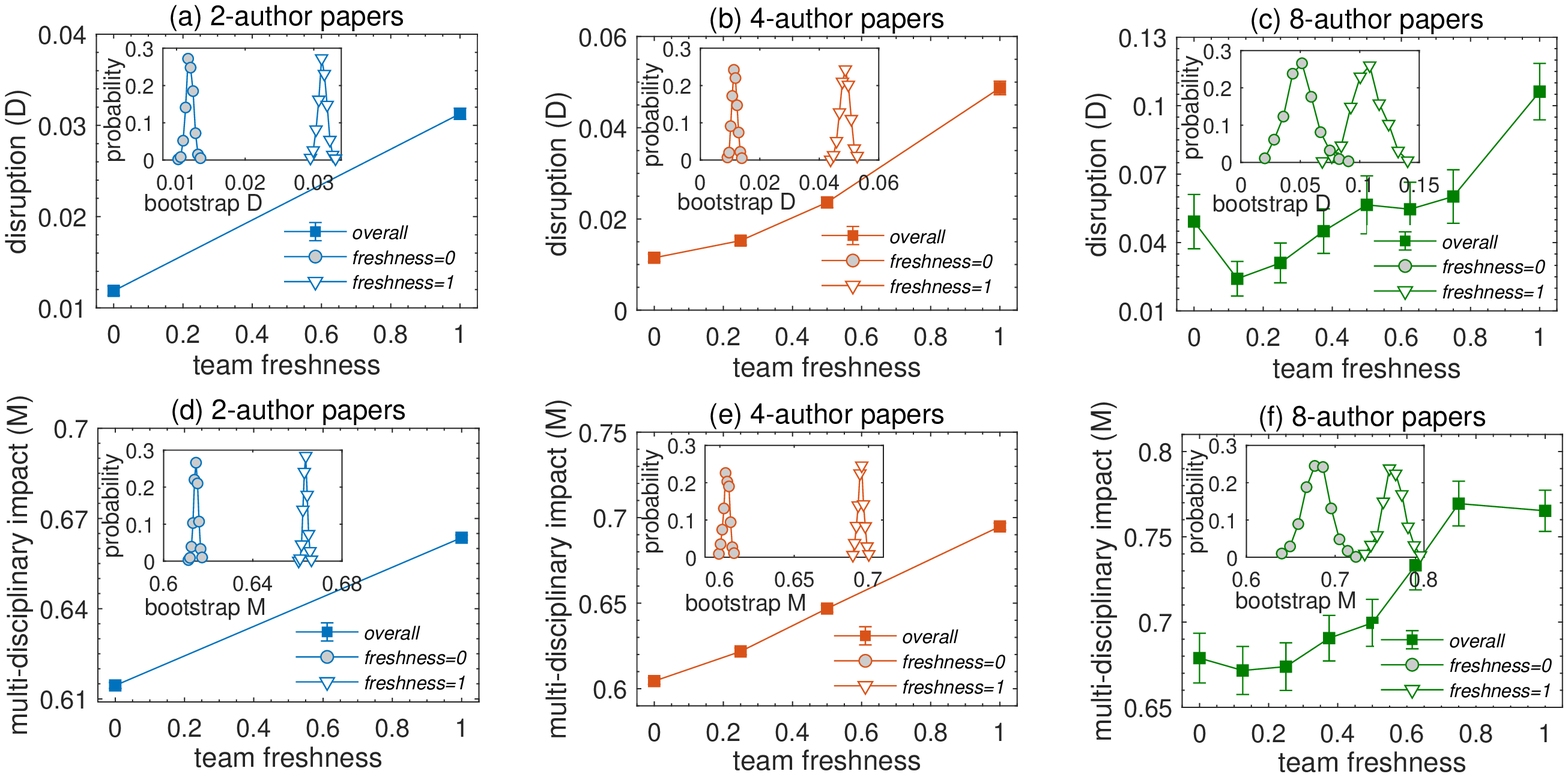}\\
  \textbf{Figure S12. Dependence of disruption (originality) and multi-disciplinarity on team freshness in computer science data.} Shown are the dependence of the disruption (originality) $D$ and multi-disciplinarity $M$ of papers on the team freshness, for (a)(d) 2-author papers, (b)(e) 4-author papers, (c)(f) 8-author papers, respectively. The results suggest that both originality and multi-disciplinarity significantly increase with team freshness. The insets show the distributions of bootstrap disruption or bootstrap multi-disciplinarity. A remarkable difference, i.e., high significance, can be observed between the distributions of $D$ of papers with team freshness 0 and 1.\label{fig2}
\end{figure}

\clearpage
\begin{figure}[h!]
  \centering
  \includegraphics[width=16.5cm]{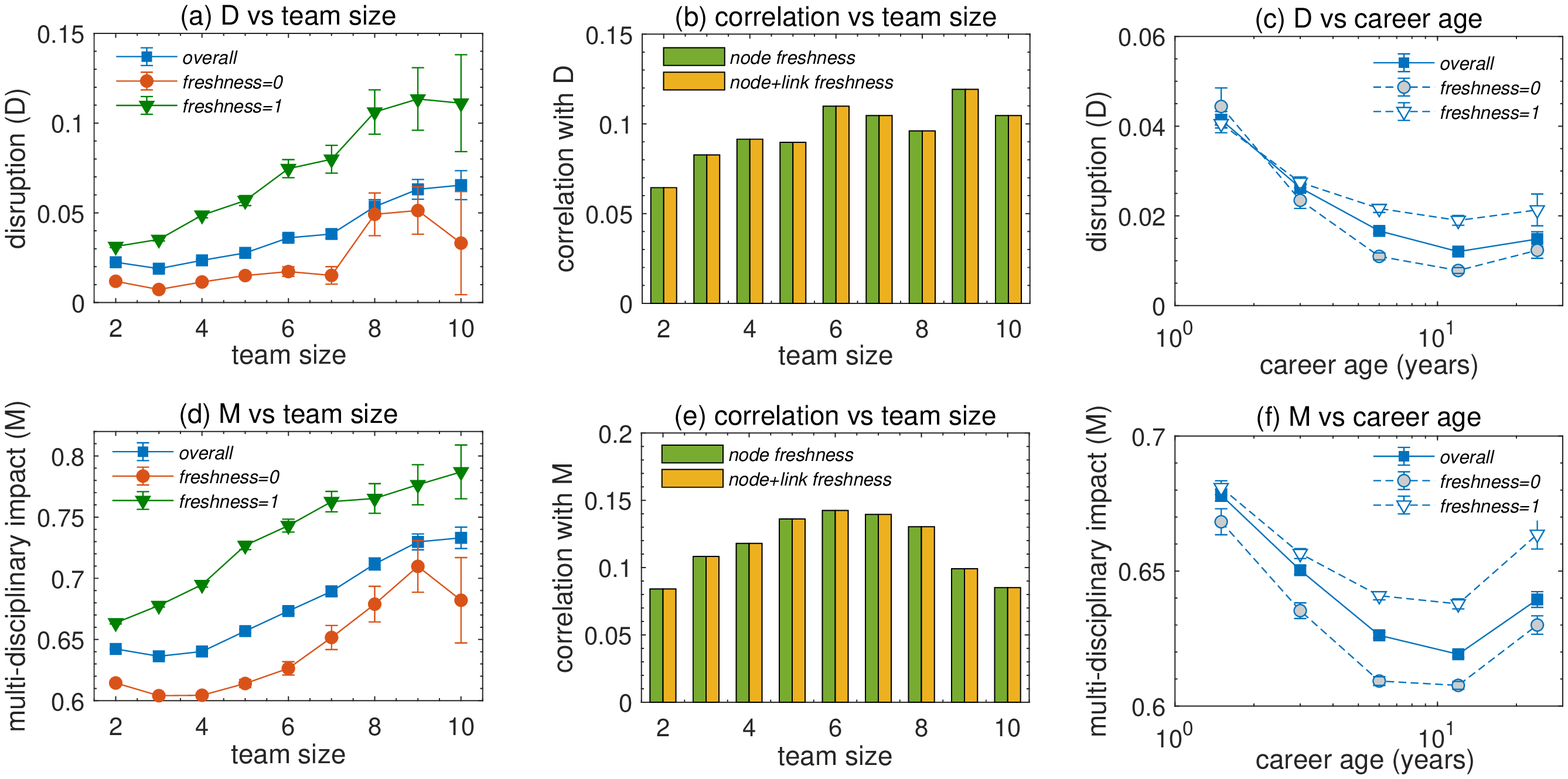}\\
  \textbf{Figure S13. Other effects of freshness in computer science data.} Plot of (a) the mean disruption (originality) $D$ and (d) the mean multi-disciplinary impact $M$ of papers versus different team sizes. For each team size, we also study the mean disruption $D$ and the mean multi-disciplinary impact $M$ of papers published by old teams (freshness=0) and fresh teams (freshness=1). (b) The Pearson correlation of node freshness and originality (disruption) for papers of different team sizes. (e) The Pearson correlation of node freshness and multi-disciplinary impact for papers of different team sizes. For comparison, we calculate the maximum Pearson correlation when we consider team freshness as a weighted linear combination of node and link freshness. We show also the dependence of (c) the mean disruption (originality) $D$ and (f) multi-disciplinarity $M$ on team members' mean career age in 2-author papers. The results suggest that both $D$ and $M$ decrease with team members' mean career age before career age$\leq10$ years.\label{fig2}
\end{figure}

\clearpage
\begin{figure}[h!]
  \centering
  \includegraphics[width=16.5cm]{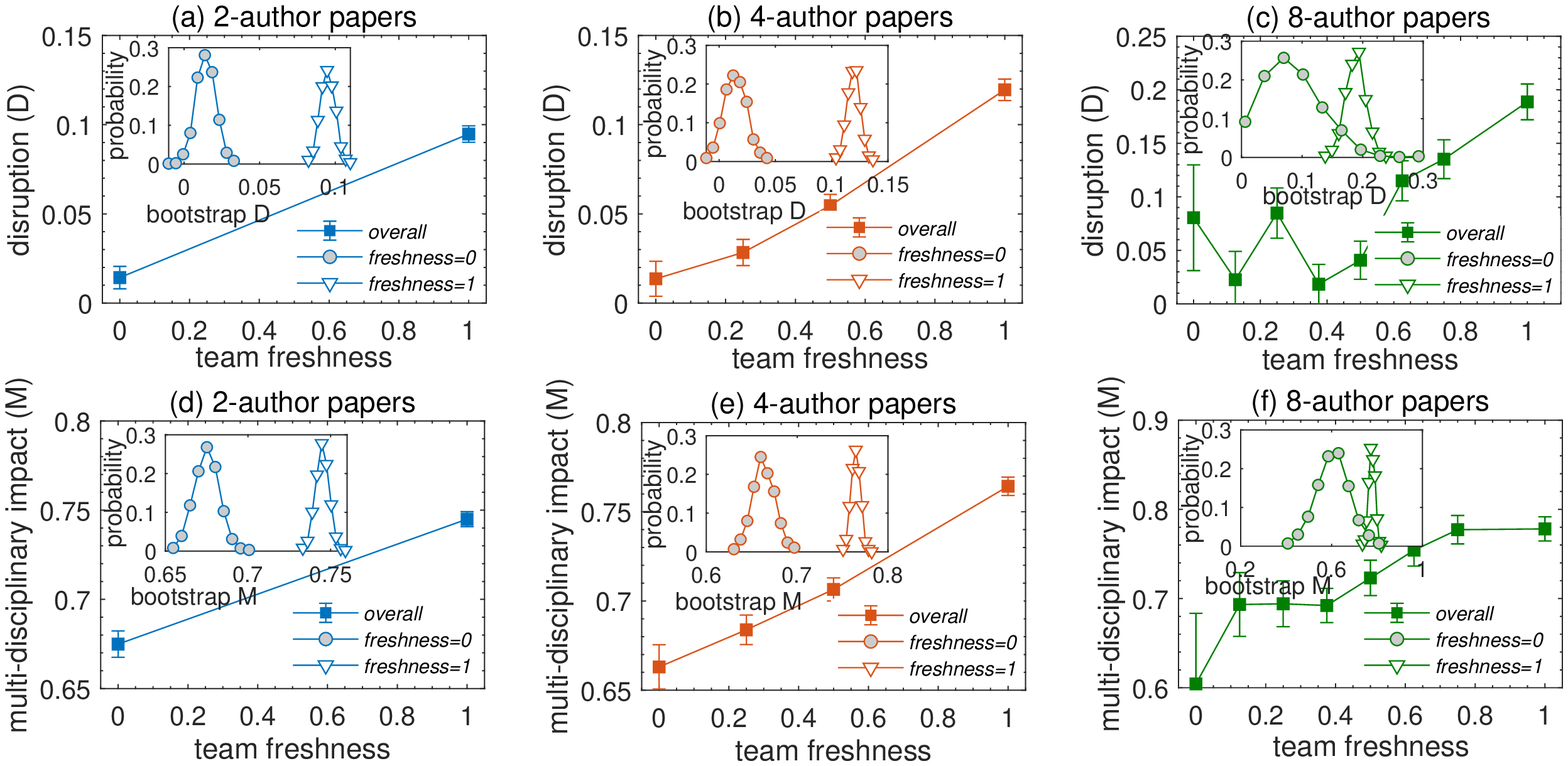}\\
  \textbf{Figure S14. Dependence of disruption (originality) and multi-disciplinarity on team freshness in Chemistry journal (JACS) data.} Shown are the dependence of the disruption (originality) $D$ and multi-disciplinarity $M$ of papers on the team freshness, for (a)(d)
2-author papers, (b)(e) 4-author papers, (c)(f) 8-author papers, respectively. The results suggest that both originality and multi-disciplinarity significantly increase with team freshness. The insets show the distributions of bootstrap disruption or bootstrap multi-disciplinarity. A remarkable difference, i.e., high significance, can be observed between the distributions of $D$ of papers with team freshness 0 and 1.\label{fig2}
\end{figure}

\clearpage
\begin{figure}[h!]
  \centering
  \includegraphics[width=16.5cm]{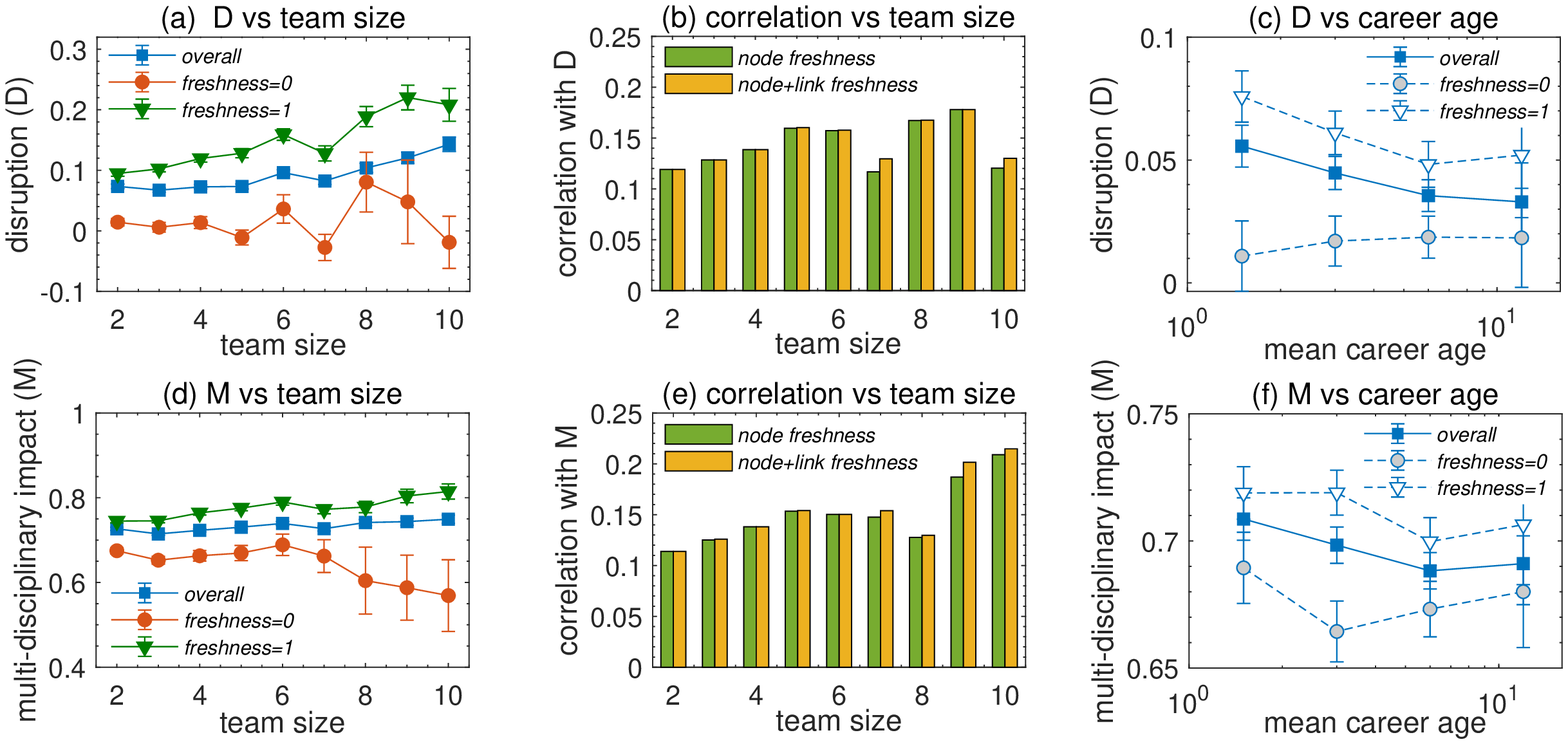}\\
  \textbf{Figure S15. Other effects of freshness in Chemistry journal (JACS) data.} Plot of (a) the mean disruption (originality) $D$ and (d) the mean multi-disciplinary impact $M$ of papers versus different team sizes. For each team size, we also study the mean disruption $D$ and the mean multi-disciplinary impact $M$ of papers published by old teams (freshness=0) and fresh teams (freshness=1). (b) The Pearson correlation of node freshness and originality (disruption) for papers of different team sizes. (e) The Pearson correlation of node freshness and multi-disciplinary impact for papers of different team sizes. For comparison, we calculate the maximum Pearson correlation when we consider team freshness as a weighted linear combination of node and link freshness. We show also the dependence of (c) the mean disruption (originality) $D$ and (f) multi-disciplinarity $M$ on team members' mean career age in 2-author papers.\label{fig2}
\end{figure}

\end{document}